\documentclass[a4paper,11pt]{article}
\pdfoutput=1 

\usepackage{jheppub} 

\usepackage[T1]{fontenc} 
\usepackage{amssymb}
\usepackage{slashbox}

\title{\begin{flushright}
\small{UAB-FT-738}
\end{flushright}
\vskip 1.5cm
\boldmath $\tau^-\to K^-\eta^{(\prime)}\nu_\tau$ decays in Chiral Perturbation Theory with Resonances}

\author[a]{R. Escribano,}
\author[a]{S. Gonz\'alez-Sol\'is,}
\author[a, 1]{P. Roig, \note{Corresponding author.}}

\affiliation[a]{Grup de F\'{\i}sica Te\`orica (Departament de F\'{\i}sica) and Institut de F\'{\i}sica d'Altes Energies (IFAE),
Universitat Aut\`onoma de Barcelona, E-08193 Bellaterra (Barcelona), Spain}

\emailAdd{rescriba@ifae.es}
\emailAdd{sgonzalez@ifae.es}
\emailAdd{proig@ifae.es}

\abstract{
We have studied the $\tau^-\to K^-\eta^{(\prime)}\nu_\tau$ decays within Chiral Perturbation Theory including resonances as explicit degrees of freedom. We have considered 
three different form factors according to treatment of final-state interactions. In increasing degree of soundness: Breit-Wigner, exponential resummation and 
dispersive representation. We find that although the first one fails in accounting for the data on the $K\eta$ mode, the other two approaches provide good fits to them which 
are sensitive to the $K^\star(1410)$ pole parameters, that are determined to be $M_{K^{\star\prime}}\,=\,\left(1330^{+27}_{-41}\right)$ MeV and $\Gamma_{K^{\star\prime}}\,=\,
\left(217^{+68}_{-122}\right)$~MeV. These values are competitive with the standard determination from $\tau^-\to(K\pi)^-\nu_\tau$ decays. The corresponding predictions for 
the $\tau^-\to K^-\eta^{\prime}\nu_\tau$ channel respect the current upper bound and hint to the discovery of this decay mode in the near future.
\vspace*{6.0cm}

PACS~: 13.35.Dx, 12.38.-t, 12.39.Fe, 11.15.Pg, 11.55.Bq
\\
\hspace*{0.5cm} Keywords~: Hadronic tau decays, Chiral Lagrangians, Dispersion relations, Analytic properties of S matrix.
}

\begin{document} 
\maketitle
\flushbottom

\section{Introduction}

Semileptonic tau decays represent a clean benchmark to study the hadronization properties of QCD due to the fact that half of the process is purely electroweak and, therefore, 
free of uncertainties at the required precision \cite{Braaten:1991qm, Braaten:1988hc, Braaten:1988ea, Braaten:1990ef, Narison:1988ni, Pich:1989pq, Davier:2005xq, Pich:2013??}. 
At the (semi-)inclusive level this allows to extract fundamental parameters of the Standard Model, most importantly the strong coupling $\alpha_S$ \cite{Davier:2008sk, 
Beneke:2008ad, Pich:2011bb, Boito:2012cr}. Tau decays containing Kaons have been split into the Cabibbo-allowed and suppressed decays \cite{Barate:1999hj, Abbiendi:2004xa} 
rendering possible determinations of the quark-mixing matrix element $|V_{us}|$ \cite{Maltman:2008ib, Antonelli:2013usa} and the mass of the strange quark \cite{Chetyrkin:1998ej, Pich:1999hc, 
Korner:2000wd, Kambor:2000dj, Chen:2001qf, Gamiz:2002nu, Gamiz:2004ar, Baikov:2004tk, Gamiz:2007qs} at high precision.\\

At the exclusive level, the largest contribution to the strange spectral function is given by the $\tau^-\to (K\pi)^-\nu_\tau$ decays ($\sim 42\%$). The corresponding differential 
decay width was measured by the ALEPH \cite{Barate:1999hj} and OPAL \cite{Abbiendi:2004xa} collaborations, and recently the B-factories BaBar \cite{Aubert:2007jh} and Belle 
\cite{Epifanov:2007rf} have published increased accuracy measurements. These high-quality data have motivated several refined studies of the related observables \cite{Jamin:2006tk, 
Moussallam:2007qc, Jamin:2008qg, Boito:2008fq, Boito:2010me} allowing for precise determinations of the $K^\star(892)$ pole parameters because this resonance gives the most 
of the contribution to the dominating vector form factor. These were also determined for the $K^\star(1410)$ resonance and the relative interference of both states was 
characterized, although with much less precision than in the case of the $K^\star(892)$ mass and width.

In order to increase the knowledge of the strange spectral function, the $\tau^-\to \left(K\pi\pi(\pi)\right)^-\nu_\tau$ decays have to be better understood (they add up to 
one third of the strange decay width), the $\tau^-\to K^- \eta \nu_\tau$ and $\tau^-\to (K\pi)^- \eta \nu_\tau$ decays being also important for that purpose. The $K^-\eta$ 
mode is also very sensitive to the $K^\star(1410)$ resonance contribution and may be competitive with the $\tau^-\to (K\pi)^-\nu_\tau$ decays in the extraction of its 
parameters. This is one of the motivations for our study of the $\tau^-\to K^- \eta^{(\prime)} \nu_\tau$ decays in this article. We will tackle the analysis of the 
$\tau^-\to \left(K\pi\right)^-\pi/\eta\,\nu_\tau$ decays along the lines employed in other three-meson \cite{GomezDumm:2003ku, Dumm:2009kj, Dumm:2009va, Dumm:2012vb} and 
one-meson radiative tau decays \cite{Guo:2010dv, Roig:2013??} elsewhere.

The $\tau^-\to K^- \eta \nu_\tau$ decays were first measured by CLEO \cite{Bartelt:1996iv} and ALEPH \cite{Buskulic:1996qs} in the '90s. Only very recently Belle \cite{Inami:2008ar} 
and BaBar \cite{delAmoSanchez:2010pc} managed to improve these measurements reducing the branching fraction to essentially half of the CLEO and ALEPH results and achieving a 
decrease of the error at the level of one order of magnitude. Belle \cite{Inami:2008ar} measured a branching ratio of $(1.58\pm0.05\pm0.09)\cdot10^{-4}$ and BaBar 
\cite{delAmoSanchez:2010pc} $(1.42\pm0.11\pm0.07)\cdot10^{-4}$, which combined to give the PDG average $(1.52\pm0.08)\cdot10^{-4}$ \cite{Beringer:1900zz}. The related decay 
$\tau^-\to K^- \eta^\prime \nu_\tau$ has not been detected yet, although an upper limit at the $90\%$ confidence level was placed by BaBar \cite{Lees:2012ks}
.

Belle's paper \cite{Inami:2008ar} cites the few existing calculations of the $\tau^-\to K^- \eta \nu_\tau$ decays based on Chiral Lagrangians \cite{Pich:1987qq, Braaten:1989zn, 
Li:1996md, Aubrecht:1981cr} and concludes that `further detailed studies of the physical dynamics in $\tau$ decays with $\eta$ mesons are required' (see also, e.g. 
Ref.~\cite{Actis:2010gg})\footnote{Very recently, the $\tau^-\to K \pi/\eta \nu_\tau$ decays have been studied \cite{Kimura:2012}. However, no satisfactory description of the 
data can be achieved in both decay channels simultaneously.}. 
Our aim is to provide a 
more elaborated analysis which takes into account the advances in this field since the publication of the quoted references more than fifteen years ago. The considered 
$\tau^-\to K^- \eta^{(\prime)} \nu_\tau$ decays are currently modeled in TAUOLA \cite{Jadach:1990mz, Jadach:1993hs}, the standard Monte Carlo generator for tau lepton decays, 
relying on phase space. We would like to provide the library with Resonance Chiral Lagrangian-based currents \cite{Shekhovtsova:2012ra, Nugent:2013hxa} that can describe well 
these decays for their analyses and for the characterization of the backgrounds they constitute to searches of rarer tau decays and new physics processes.

Our paper is organized as follows: the hadronic matrix element and the participating vector and scalar form factors are defined in section \ref{M.e. and decay width}, where 
the differential decay distribution in terms of the latter is also given. These form factors are derived within Chiral Perturbation Theory ($\chi PT$) \cite{Weinberg:1978kz, 
Gasser:1983yg, Gasser:1984gg} including resonances ($R\chi T$) \cite{Ecker:1988te, Ecker:1989yg} in section \ref{FFs}. Three different options according to treatment of 
final-state interactions in these form factors are discussed in section \ref{FSI} and will be used in the remainder of the paper. In section \ref{Pred Keta}, the 
$\tau^-\to K^-\eta\nu_\tau$ decay observables are predicted based on the knowledge of the $\tau^-\to (K\pi)^-\nu_\tau$ decays. These results are then improved in section 
\ref{Fit Keta} by fitting the BaBar and Belle $\tau^-\to K^-\eta\nu_\tau$ data. We provide our predictions on the $\tau^-\to K^-\eta^\prime\nu_\tau$ decays in section 
\ref{Pred Ketap} and present our conclusions in section \ref{Concl}.

\section{Matrix elements and decay width}\label{M.e. and decay width}
We fix our conventions from the general parametrization of the scalar and vector $K^+\eta^{(\prime)}$ matrix elements \cite{Gasser:1984ux}:
\begin{equation}\label{general definition}
 \left\langle\eta^{(\prime)} \Big|\bar{s}\gamma^\mu u\Big| K^+\right\rangle=c^V_{K\eta^{(\prime)}}\left[\left(p_{\eta^{(\prime)}}+p_K\right)^\mu f_+^{K^+ \eta^{(\prime)}}(t)
+(p_K-p_{\eta^{(\prime)}})^\mu f_-^{K ^+\eta^{(\prime)}}(t)\right]\,,
\end{equation}
where $t=(p_K-p_{\eta^{(\prime)}})^2$. From eq. (\ref{general definition}) one has
\begin{equation}\label{m.e. fmas fmenos}
 \left\langle K^-\eta^{(\prime)} \Big|\bar{s}\gamma^\mu u\Big| 0\right\rangle=c^V_{K\eta^{(\prime)}}\left[\left(p_{\eta^{(\prime)}}-p_K\right)^\mu f_+^{K^- \eta^{(\prime)}}(s)
-q^\mu f_-^{K ^-\eta^{(\prime)}}(s)\right]\,,
\end{equation}
with $q^\mu=\left(p_{\eta^{(\prime)}}+p_K\right)^\mu$, $s=q^2$ and $c^V_{K\eta^{(\prime)}}=-\sqrt{\frac{3}{2}}$. Instead of $f_-^{K^- \eta^{(\prime)}}(s)$ one can use 
$f_0^{K^- \eta^{(\prime)}}(s)$ defined through
\begin{equation}\label{definition f0}
 \left\langle 0 \Big|\partial_\mu(\bar{s}\gamma^\mu u)\Big| K^-\eta^{(\prime)}\right\rangle=i(m_s-m_u)\left\langle 0 \Big|\bar{s}u\Big| K^-\eta^{(\prime)}\right\rangle\equiv i\Delta_{K\pi}c^S_{K^-\eta^{(\prime)}}f_0^{K ^-\eta^{(\prime)}}(s)\,,
\end{equation}
with
\begin{equation}
c^S_{K^-\eta}= -\frac{1}{\sqrt{6}}\,,\quad c^S_{K^-\eta^\prime}= \frac{2}{\sqrt{3}}\,,\quad \Delta_{PQ}=m_P^2-m_Q^2\,. 
\end{equation}
The mass renormalization $m_s-\bar{m}$ in $\chi PT$ (or $R\chi T$)
needs to be taken into account to define $f_0^{K ^-\eta^{(\prime)}}(s)$ and $\bar{m}=(m_d+m_u)/2$ has been introduced. 
We will take $\Delta_{K\pi}\Big|^{QCD}=\Delta_{K\pi}$, which is an excellent approximation. From eqs.~(\ref{m.e. fmas fmenos}) and (\ref{definition f0}) one gets
\begin{equation}\label{Had m.e.}
 \left\langle K^-\eta^{(\prime)} \Big|\bar{s}\gamma^\mu u\Big| 0\right\rangle=\left[\left(p_{\eta^{(\prime)}}-p_K\right)^\mu +\frac{\Delta_{K \eta^{(\prime)}}}{s}q^\mu\right]c^V_{K^-\eta^{(\prime)}}f_+^{K^-\eta^{(\prime)}}(s)+
\frac{\Delta_{K \pi}}{s}q^\mu c^S_{K^-\eta^{(\prime)}} f_0^{K^-\eta^{(\prime)}}(s)\,,
\end{equation}
and the normalization condition
\begin{equation}\label{condition origin}
 f_+^{K^-\eta^{(\prime)}}(0)=-\frac{c^S_{K^-\eta^{(\prime)}}}{c^V_{K^-\eta^{(\prime)}}}\frac{\Delta_{K\pi}}{\Delta_{K\eta^{(\prime)}}}f_0^{K^-\eta^{(\prime)}}(0)\,,
\end{equation}
which is obtained from
\begin{equation}
 f_-^{K ^-\eta^{(\prime)}}(s)=-\frac{\Delta_{K\eta^{(\prime)}}}{s}\left[\frac{c^S_{K^-\eta^{(\prime)}}}{c^V_{K^-\eta^{(\prime)}}}\frac{\Delta_{K\pi}}{\Delta_{K\eta^{(\prime)}}}f_0^{K^-\eta^{(\prime)}}(s)+f_+^{K^-\eta^{(\prime)}}(s)\right]\,.
\end{equation}

In terms of these form factors, the differential decay width reads
\begin{eqnarray} \label{spectral function}
& & \frac{d\Gamma\left(\tau^-\to K^-\eta^{(\prime)}\nu_\tau\right)}{d\sqrt{s}} = \frac{G_F^2M_\tau^3}{32\pi^3s}S_{EW}\Big|V_{us}f_+^{K^-\eta^{(\prime)}}(0)\Big|^2
\left(1-\frac{s}{M_\tau^2}\right)^2\\
& & \left\lbrace\left(1+\frac{2s}{M_\tau^2}\right)q_{K\eta^{(\prime)}}^3(s)\Big|\widetilde{f}_+^{K^-\eta^{(\prime)}}(s)\Big|^2+\frac{3\Delta_{K\eta^{(\prime)}}^2}{4s}q_{K\eta^{(\prime)}}(s)\Big|\widetilde{f}_0^{K^-\eta^{(\prime)}}(s)\Big|^2\right\rbrace\,,\nonumber
\end{eqnarray}
where
\begin{eqnarray}\label{definitions}
& &  q_{PQ}(s)=\frac{\sqrt{s^2-2s\Sigma_{PQ}+\Delta_{PQ}^2}}{2\sqrt{s}}\,,\quad \sigma_{PQ}(s)=\frac{2q_{PQ}(s)}{\sqrt{s}}\theta\left(s-(m_P+m_Q)^2\right)\,,\nonumber\\
& & \Sigma_{PQ}=m_P^2+m_Q^2\,
,\quad \widetilde{f}_{+,0}^{K^-\eta^{(\prime)}}(s)=\frac{f_{+,0}^{K^-\eta^{(\prime)}}(s)}{f_{+,0}^{K^-\eta^{(\prime)}}(0)}\,,
\end{eqnarray}
and $S_{EW} = 1.0201$ \cite{Erler:2002mv} represents an electro-weak correction factor.

We have considered the $\eta-\eta^\prime$ mixing up to next-to-leading order in the combined expansion in $p^2$, $m_q$ and $1/N_C$ \cite{Kaiser:1998ds, Kaiser:2000gs} 
(see the next section for the introduction of the large-$N_C$ limit of QCD \cite{'tHooft:1973jz, 'tHooft:1974hx, Witten:1979kh} applied to the light-flavoured mesons). 
In this way it is found that $\Big|V_{us}f_+^{K^-\eta}(0)\Big|=\Big|V_{us}f_+^{K^-\pi^0}(0)\mathrm{cos}\theta_P\Big|$, $\Big|V_{us}f_+^{K^-\eta^\prime}(0)\Big|=\Big|V_{us}f_+^{K^-\pi^0}(0)\mathrm{sin}\theta_P\Big|$, 
where 
$\theta_P=(-13.3\pm1.0)^\circ$ \cite{Ambrosino:2006gk}.

The best access to $\Big|V_{us}f_+^{K^-\pi^0}(0)\Big|$ is through semi-leptonic Kaon decay data. We will use the value $0.21664\pm0.00048$ \cite{Beringer:1900zz, Antonelli:2010yf}. 
Eq.~(\ref{spectral function}) makes manifest that the unknown strong-interaction dynamics is encoded in the tilded form factors, $\widetilde{f}_{+,0}^{K^-\eta^{(\prime)}}(s)$ 
which will be subject of our analysis in the following section. We will see in particular that the use of $\widetilde{f}_{+,0}^{K^-\eta^{(\prime)}}(s)$ instead of the 
untilded form factors yields more compact expressions that are symmetric under the exchange $\eta\leftrightarrow\eta^\prime$, see eqs.(\ref{RChT VFFs}) and (\ref{RChT SFFs def}).

\section{Scalar and vector form factors in $\boldsymbol{\chi PT}$ with resonances}\label{FFs}
Although there is no analytic method to derive the $\widetilde{f}_{+,0}^{K^-\eta^{(\prime)}}(s)$ form factors directly from the QCD Lagrangian, its symmetries are nevertheless 
useful to reduce the model dependence to a minimum and keep as many properties of the fundamental theory as possible.

$\chi PT$ \cite{Weinberg:1978kz, Gasser:1983yg, Gasser:1984gg}, the effective field theory of QCD at low energies, is built as an expansion in even powers of the ratio between 
the momenta or masses of the lightest pseudoscalar mesons over the chiral symmetry breaking scale, which is of the order of one GeV. As one approaches the energy region where 
new degrees of freedom -the lightest meson resonances- become active, $\chi PT$ ceases to provide a good description of the Physics (even including higher-order corrections 
\cite{Bijnens:1999sh, Bijnens:1999hw, Bijnens:2001bb}) and these resonances must be incorporated to the action of the theory. This is done without any ad-hoc dynamical 
assumption by $R\chi T$ in the convenient antisymmetric tensor formalism that avoids the introduction of local $\chi PT$ terms at next-to-leading order in the chiral expansion 
since their contribution is recovered upon integrating the resonances out \cite{Ecker:1988te, Ecker:1989yg}. The building of the Resonance Chiral Lagrangians is driven by the 
spontaneous symmetry breakdown of QCD realized in the meson sector, the discrete symmetries of the strong interaction and unitary symmetry for the resonance multiplets. The 
expansion parameter of the theory is the inverse of the number of colours of the gauge group, $1/N_C$. Despite $N_C$ not being small in the real world, the fact that phenomenology 
supports this approach to QCD \cite{Manohar:1998xv, Pich:2002xy} hints that the associated coefficients of the expansion are small enough to warrant a meaningful perturbative 
approach based on it. At leading order in this expansion there is an infinite number of radial excitations for each resonance with otherwise the same quantum numbers that are 
strictly stable and interact through local effective vertices only at tree level.

The relevant effective Lagrangian for the lightest resonance nonets reads~\footnote{We comment on its extension to the infinite spectrum predicted in the $N_C\to\infty$ limit 
in the paragraph below eq.~(\ref{JOP FFs}).}:
\begin{eqnarray}
\label{eq:ret} {\cal L}_{\rm R\chi T}   & \doteq   & {\cal L}_{\rm kin}^{\rm V,S}\, + \, \frac{F^2}{4}\langle u_{\mu} u^{\mu} + \chi _+\rangle \, + \, 
\frac{F_V}{2\sqrt{2}} \langle V_{\mu\nu} f_+^{\mu\nu}\rangle\,+\,i \,\frac{G_V}{\sqrt{2}} \langle V_{\mu\nu} u^\mu u^\nu\rangle \,+\, c_d \langle S u_{\mu} u^{\mu}\rangle 
\,+\, c_m \langle S \chi_ +\rangle\,,\nonumber\\
\label{lagrangian}
\end{eqnarray}
where all coupling constants are real, $F$ is the pion decay constant and we follow the conventions of Ref.~\cite{Ecker:1988te}. Accordingly, $\langle \rangle$ stands for 
trace in flavour space, and $u^\mu$, $\chi_+$ and $f_+^{\mu\nu}$ are defined by
\begin{eqnarray}
u^\mu & = & i\,u^\dagger\, D^\mu U\, u^\dagger \,,\nonumber \\
\chi_\pm & = & u^\dagger\, \chi \, u^\dagger\, \pm u\,\chi^\dagger\, u\,, \nonumber \\
f_\pm^{\mu\nu} & = & u^\dagger\, F_L^{\mu\nu}\, u^\dagger\, \pm
u\,F_R^{\mu\nu}\, u \ ,
\end{eqnarray}
where $u$ ($U=u^2$), $\chi$ and $F_{L,R}^{\mu\nu}$ are $3\times 3$ matrices that contain light pseudoscalar fields, current quark masses and external left and right currents, 
respectively. The matrix $V^{\mu\nu}$ ($S$) includes the lightest vector (scalar) meson multiplet \footnote{In the $N_C\to\infty$ limit of QCD the lightest scalar meson 
multiplet does not correspond to the one including the $f_0(600)$ (or $\sigma$ meson) \cite{Cirigliano:2003yq}, but rather to the one including the $f_0(1370)$ resonance.}, and ${\cal L}_{\rm kin}^{\rm V,S}$ stands for these resonances kinetic term. 
We note that resonances with other quantum numbers do not contribute to the considered processes (like the axial-vector and pseudoscalar resonances, which have the wrong parity). 

The computation of the vector form factors yields
\begin{equation} \label{RChT VFFs}
 \tilde{f}_+^{K^-\eta}(s)=\frac{f_+^{K^-\eta}(s)}{f_+^{K^-\eta}(0)}=1+\frac{F_V G_V}{F^2}\frac{s}{M_{K^\star}^2-s}\,=\frac{f_+^{K^-\eta^\prime}(s)}{f_+^{K^-\eta^\prime}(0)}=\, \tilde{f}_+^{K^-\eta^\prime}(s)\,,
\end{equation}
because $f_+^{K^-\eta}(0)=\cos\theta_{P}$ and $f_+^{K^-\eta^\prime}(0)=\sin\theta_{P}$. We recall that the normalization of the $K\pi$ vector form factor, $f_+^{K^-\pi}(0)$, 
was pre-factored in eq.~(\ref{spectral function}) together with $|V_{us}|$.

The strangeness changing scalar form factors and associated S-wave scattering within $R\chi T$ have been investigated in a series of papers by Jamin, Oller and Pich 
\cite{Jamin:2000wn, Jamin:2001zq, Jamin:2006tk,Jamin:2006tj} (see also Ref.~\cite{Bernard:1990kw}). The computation of the scalar form factors gives:
\begin{eqnarray}\label{RChT SFFs}
&& \tilde{f}_0^{K^-\eta}(s)=\frac{f_0^{K^-\eta}(s)}{f_0^{K^-\eta}(0)}=\frac{1}{f_0^{K^-\eta}(0)}\left[\cos\theta_{P}f_0^{K^-\eta_8}(s)\Big|_{\eta_8\to\eta}+2\sqrt{2}\mathrm{sin}\theta_Pf_0^{K^-\eta_1}(s)\Big|_{\eta_1\to\eta}\right]\,,\;\;\;\;\;\;\;\;\\
&& \tilde{f}_0^{K^-\eta^\prime}(s)=\frac{f_0^{K^-\eta^\prime}(s)}{f_0^{K^-\eta^\prime}(0)}=\frac{1}{f_0^{K^-\eta^\prime}(0)}\left[\mathrm{cos}\theta_Pf_0^{K^-\eta_1}(s)\Big|_{\eta_1\to\eta^\prime}-\frac{1}{2\sqrt{2}}\mathrm{sin}\theta_Pf_0^{K^-\eta_8}(s)\Big|_{\eta_8\to\eta^\prime}\right]\,,\nonumber
\end{eqnarray}
and can be written in terms of the $f_0^{K^-\eta_8}(s)$, $f_0^{K^-\eta_1}(s)$ form factors computed in Ref.\cite{Jamin:2001zq}:
\begin{eqnarray}\label{JOP FFs}
 f_0^{K^-\eta_8}(s) & = & 1+\frac{4 c_m}{F^2(M_S^2-s)}\left[c_d(s-m_K^2-p_{\eta_8}^2)+c_m(5m_K^2-3m_\pi^2)\right]+\frac{4c_m(c_m-c_d)}{F^2M_S^2}(3m_K^2-5m_\pi^2)\,,\nonumber\\
 f_0^{K^-\eta_1}(s) & = & 1+\frac{4c_m}{F^2(M_S^2-s)}\left[c_d(s-m_K^2-p_{\eta_1}^2)+c_m2m_K^2\right]-\frac{4c_m(c_m-c_d)}{F^2 M_S^2}2m_\pi^2\,,
\end{eqnarray}
where, for the considered flavour indices, $S$ should correspond to the $K^\star_0(1430)$ resonance. Besides $f_0^{K^-\pi}(0)=f_+^{K^-\pi}(0)$ (see the comment below 
equation~(\ref{RChT VFFs})) it has also been used that
\begin{eqnarray}
 f_0^{K^-\eta}(0) & = & \cos\theta_{P}\left(1+\frac{\Delta_{K\eta}+3\Delta_{K\pi}}{M_S^2}\right)+2\sqrt{2}\sin\theta_P\left(1+\frac{\Delta_{K\eta}}{M_S^2}\right)\,,\nonumber\\
 f_0^{K^-\eta^\prime}(0) & = & \cos\theta_{P}\left(1+\frac{\Delta_{K\eta}}{M_S^2}\right)+\sin\theta_P\left(1+\frac{\Delta_{K\eta}+3\Delta_{K\pi}}{M_S^2}\right)\,.
\end{eqnarray}
Indeed, using our conventions, the tilded scalar form factors become simply
\begin{equation}\label{RChT SFFs def}
  \tilde{f}_0^{K^-\eta}(s)=\frac{f_0^{K^-\eta}(s)}{f_0^{K^-\eta}(0)}=1+\frac{c_d c_m}{4 F^2}\frac{s}{M_S^2-s}\,=\frac{f_+^{K^-\eta^\prime}(s)}{f_0^{K^-\eta^\prime}(0)}=\, \tilde{f}_0^{K^-\eta^\prime}(s)\,,
\end{equation}
that is more compact than eqs.~(\ref{RChT SFFs}), (\ref{JOP FFs}) and displays the same symmetry $\eta\leftrightarrow\eta^\prime$ than the vector form factors in 
eq.~(\ref{RChT VFFs}).

The computation of the leading order amplitudes in the large-$N_C$ limit within $R\chi T$ demands, however, the inclusion of an infinite tower of resonances per set of 
quantum numbers \footnote{We point out that there is no limitation in the $R\chi T$ Lagrangians in this respect. In particular, a second multiplet of resonances has been 
introduced in the literature \cite{SanzCillero:2002bs, Mateu:2007tr} and bi- and tri-linear operators in resonance fields have been used \cite{GomezDumm:2003ku, 
RuizFemenia:2003hm, Cirigliano:2004ue, Cirigliano:2005xn, Cirigliano:2006hb, Kampf:2011ty}.}. Although the masses of the large-$N_C$ states depart slightly from the 
actually measured particles \cite{Masjuan:2007ay} only the second vector state, i.e. the $K^\star(1410)$ resonance, will have some impact on the considered decays. Accordingly, 
we will replace the vector form factor in eq.~(\ref{RChT VFFs}) by
\begin{equation} \label{RChT VFFs2Res}
 \tilde{f}_+^{K^-\eta^{(\prime)}}(s)=1+\frac{F_V G_V}{F^2}\frac{s}{M_{K^\star}^2-s}+\frac{F_V^\prime G_V^\prime}{F^2}\frac{s}{M_{K^\star\prime}^2-s}\,,
\end{equation}
where the operators with couplings $F_V^\prime$ and $G_V^\prime$ are defined in analogy with the corresponding unprimed couplings in eq.~(\ref{lagrangian}).

If we require that the $f_+^{K^-\eta^{(\prime)}}(s)$ and $f_0^{K^-\eta^{(\prime)}}(s)$ form factors vanish for $s\to\infty$ at least as $1/s$ \cite{Lepage:1979zb, 
Lepage:1980fj}, we obtain the short-distance constraints
\begin{equation}\label{shortdistance}
 F_V G_V + F_V^\prime G_V^\prime= F^2\,,\quad 4 c_d c_m = F^2\,,\quad c_d-c_m=0\,,
\end{equation}
which yield the form factors
\begin{eqnarray} \label{FFs with short distance constraints}
& & \tilde{f}_+^{K^-\eta}(s)=\frac{M_{K^\star}^2+\gamma s}{M_{K^\star}^2-s}-\frac{\gamma s}{M_{K^\star\prime}^2-s}= \tilde{f}_+^{K^-\eta^\prime}(s)\,,\\
& & \tilde{f}_0^{K^-\eta}(s) = \frac{M_S^2}{M_S^2-s}\,=\,\tilde{f}_0^{K^-\eta^\prime}(s)\,,\nonumber
\end{eqnarray}
where $\gamma=-\frac{F_V^\prime G_V^\prime}{F^2}=\frac{F_VG_V}{F^2}-1$ \cite{Jamin:2006tk, Jamin:2008qg, Boito:2008fq, Boito:2010me}. We note that we are disregarding the 
modifications introduced by the heavier resonance states to the relation (\ref{shortdistance}) and to the definition of $\gamma$.
\section{Different form factors according to treatment of final-state interactions}\label{FSI}
The form factors in eqs.(\ref{FFs with short distance constraints}) diverge when the exchanged resonance is on-mass shell and, consequently, cannot represent the underlying 
dynamics that may peak in the resonance region but does not certainly show a singular behaviour. This is solved by considering a next-to-leading order effect in the large-$N_C$ 
counting, as it is a non-vanishing resonance width \footnote{Other corrections at this order are neglected. Phenomenology seems to support that this is the predominant 
contribution.}. Moreover, since the participating resonances are not narrow, an energy-dependent width needs to be considered. A precise formalism-independent definition of 
the off-shell vector resonance width within $R\chi T$ has been given in Ref.~\cite{GomezDumm:2000fz} and employed successfully in a variety of phenomenological studies. Its 
application to the $K^*(892)$ resonance gives
\begin{eqnarray}\label{K^* width predicted}
 \Gamma_{K^*}(s) & = & \frac{G_V^2 M_{K^*} s}{64 \pi F^4}
\bigg[\sigma_{K\pi}^3(s)+ \mathrm{cos}^2\theta_P \sigma_{K\eta}^3(s) + \mathrm{sin}^2\theta_P\sigma_{K\eta^\prime}^3(s)\bigg] \,,
\end{eqnarray}
where $\sigma_{PQ}(s)$ was defined in eq.~(\ref{definitions}). Several analyses of the $\pi\pi$ \cite{SanzCillero:2002bs, Pich:2001pj, Dumm:2013zh} and $K\pi$ \cite{Jamin:2008qg, 
Boito:2008fq, Boito:2010me} form factors where the $\rho(770)$ and $K^\star(892)$ prevail respectively, have probed the energy-dependent width of these resonances with precision. 
 Although the predicted width \cite{Guerrero:1997ku} turns to be quite accurate, it is not optimal to achieve a very precise description of the data and, instead, it is 
better to allow (as we will do in the remainder of the paper) the on-shell width to be a free parameter and write
\begin{eqnarray}\label{K^* width}
 \Gamma_{K^*}(s) & = & \Gamma_{K^*}\frac{s}{M_{K^*}^2}\frac{\sigma_{K\pi}^3(s)+\mathrm{cos}^2\theta_P \sigma_{K\eta}^3(s) + \mathrm{sin}^2\theta_P\sigma_{K\eta^\prime}^3(s)}{\sigma_{K\pi}^3(M_{K^*}^2)}\,,
\end{eqnarray}
where it has been taken into account that at the $M_{K^*}$-scale the only absorptive cut is given by the elastic contribution.

In the case of the $K^\star(1410)$ resonance there is no warranty that the $KP$ ($P=\pi$, $\eta$, $\eta^\prime$) cuts contribute in the proportion given in eqs.(\ref{K^* width predicted}) 
and (\ref{K^* width}). We will assume that the lightest $K\pi$ cut dominates and use throughout that
\begin{equation}\label{Kstarprimewidth}
 \Gamma_{K^{\star\prime}}(s)\,=\,\Gamma_{K^{\star\prime}}\frac{s}{M_{K^{\star\prime}}^2}\frac{\sigma_{K\pi}^3(s)}{\sigma_{K\pi}^3(M_{K^{\star\prime}}^2)}\,.
\end{equation}
The scalar resonance width can also be computed in $R\chi T$ similarly \cite{Ecker:1988te, GomezDumm:2000fz}. In the case of the $K^\star_0(1430)$ it reads
\begin{equation}\label{Gamma S computed}
 \Gamma_{S}(s)\,=\,\Gamma_{S_0}\left(M_S^2\right)\left(\frac{s}{M_S^2}\right)^{3/2}\frac{g(s)}{g\left(M_S^2\right)}\,,
\end{equation}
with
\begin{eqnarray}\label{g(s)}
 g(s) & = & \frac{3}{2}\sigma_{K\pi}(s)+\frac{1}{6}\sigma_{K\eta}(s)\left[\mathrm{cos}\theta_P\left(1+\frac{3\Delta_{K\pi}+\Delta_{K\eta}}{s}\right)+2\sqrt{2}\mathrm{sin}\theta_P\left(1+\frac{\Delta_{K\eta}}{s}\right)\right]^2\nonumber\\
& & +\frac{4}{3}\sigma_{K\eta^\prime}(s)\left[\mathrm{cos}\theta_P\left(1+\frac{\Delta_{K\eta^\prime}}{s}\right)-\frac{\mathrm{sin}\theta_P}{2\sqrt{2}}\left(1+\frac{3\Delta_{K\pi}+\Delta_{K\eta^\prime}}{s}\right)\right]^2\,.
\end{eqnarray}

At this point, different options for the inclusion of the resonances width arise. The most simple prescription is to replace $M_R^2-s$ by $M_R^2-s-iM_R\Gamma_R(s)$ in 
eqs.~(\ref{FFs with short distance constraints}). We shall call this option `dipole model', or simply `Breit-Wigner (BW) model'. One should pay attention to the fact that 
analyticity of a quantum field theory imposes certain relations between the real and imaginary parts of the amplitudes. In particular, there is one between the real and 
imaginary part of the relevant two-point function. At the one-loop level its imaginary part is proportional to the meson width but the real part (which is neglected in this 
model) is non-vanishing. As a result, the Breit-Wigner treatment breaks analyticity at the leading non-trivial order.

Instead, one can try to devise a mechanism that keeps the complete complex two-point function. Ref.~\cite{Guerrero:1997ku} used an Omn\`es resummation of final-state 
interactions in the vector form factor that was consistent with analyticity at next-to-leading order. The associated violations were small and consequently neglected in their 
study of the $\pi\pi$ observables. This strategy was also exported to the $K\pi$ decays of the $\tau$ in Refs.~\cite{Jamin:2006tk, Jamin:2008qg} where it yielded 
remarkable agreement with the data. We will call this approach to the vector form factor `the exponential parametrization' (since it exponentiates the real part of the 
relevant loop function) and refer to it by the initials of the authors who studied the $K\pi$ system along these lines, `JPP'.

A decade after, a construction that ensures analyticity of the vector form factor exactly was put forward in Ref.~\cite{Boito:2008fq} and applied successfully to the study of 
the $K\pi$ tau decays. It is a dispersive representation of the form factor where the input phaseshift, which resums the whole loop function in the denominator of 
eq.~(\ref{FFs with short distance constraints}), is proportional to the ratio of the imaginary and real parts of this form factor. This method also succeeded in its 
application to the di-pion system \cite{Dumm:2013zh}, where it was rephrased in a way which makes chiral symmetry manifest at next-to-leading order. We will name this method 
`dispersive representation' or `BEJ', by the authors who pioneered it in the $K\pi$ system.

We would like to stress that the Breit-Wigner model is consistent with $\chi PT$ only at leading order, while the exponential parametrization (JPP) and the dispersive 
representation (BEJ) reproduce the chiral limit results up to next-to-leading order and including the dominant contributions at the next order \cite{Guerrero:1998hd}.

In the dispersive approach to the study of the di-pion and Kaon-pion systems it was possible to achieve a unitary description in the elastic region that could be extended up 
to $s_{inel}=4m_K^2$ (the $4\pi$ cut, which is phase-space and large-$N_C$ suppressed is safely neglected) and $s_{inel}=(m_K+m_\eta)^2$, respectively. Most devoted studies 
of these form factors neglect -in one way or another- inelasticities and coupled-channel effects beyond $s_{inel}$ in them \footnote{See, however, Ref.~\cite{Moussallam:2007qc}, 
which includes coupled channels for the $K\pi$ vector form factor.}, an approximation that seems to be supported by the impressive agreement with the data sought. However, 
this overlook of the problem seems to be questionable in the case of the $\tau^-\to K^-\eta^{(\prime)}\nu_\tau$ decays where we are concerned with the first (second) 
inelastic cuts.

An advisable solution may come from the technology developed for the scalar form factors that were analyzed in a coupled channel approach in Refs.~\cite{Jamin:2001zq, Jamin:2001zr, 
Jamin:2006tj} (for the strangeness-changing form factors) \footnote{We will use these unitarized scalar form factors instead of the one in 
eq.~(\ref{FFs with short distance constraints}) in the JPP and BEJ treatments (see above).} and \cite{Guo:2012ym, Guo:2012yt} (for the strangeness-conserving ones) unitarizing 
$SU(3)$ and $U(3)$ (respectively) $\chi PT$ with explicit exchange of resonances \cite{Guo:2011pa}. However, given the large errors of the $\tau^-\to K^-\eta\nu_\tau$ decay 
spectra measured by the BaBar \cite{delAmoSanchez:2010pc} and Belle \cite{Inami:2008ar} Collaborations and the absence of data on the $K^-\eta^{\prime}$ channel we consider 
that it is not timely to perform such a cumbersome numerical analysis in the absence of enough experimental guidance \footnote{One could complement this poorly known sector 
with the information from meson-meson scattering on the relevant channels \cite{GomezNicola:2001as}. Our research at next-to-leading order in the $1/N_C$ expansion treating 
consistently the $\eta-\eta^\prime$ mixing \cite{Kaiser:1998ds, Kaiser:2000gs, Escribano:2010wt} is in progress.}. For this reason we have attempted 
to obviate the inherent inelasticity of the $K\eta^{(\prime)}$ channels and tried an elastic description, where the form factor that defines the input phaseshift is given by 
eq.~(\ref{FFs with short distance constraints}) with $\Gamma_{K^\star}(s)$ defined analogously to $\Gamma_{K^{\star\prime}}(s)$, i.e., neglecting the inelastic cuts. We 
anticipate that the accord with data supports this procedure until more precise measurements demand a better approximation.

Let us recapitulate the different alternatives for the treatment of final-state interactions that will be employed in sections \ref{Pred Keta}-\ref{Pred Ketap} to study the 
$\tau^-\to K^-\eta^{(\prime)}\nu_\tau$ decays. The relevant form factors will be obtained from eqs.(\ref{FFs with short distance constraints}) in each case by:
\begin{itemize}
 \item Dipole model (Breit-Wigner): $M_R^2-s$ will be replaced by $M_R^2-s-iM_R\Gamma_R(s)$ with $\Gamma_{K^\star}(s)$ and $\Gamma_S(s)$ given by eqs. (\ref{K^* width}) and (\ref{Gamma S computed}).
 \item Exponential parametrization (JPP): The Breit-Wigner vector form factor described above is multiplied by the exponential of the real part of the loop function. The 
unitarized scalar form factor \cite{Jamin:2001zq} will be employed. The relevant formulae can be found in appendix \ref{app}.
 \item Dispersive representation (BEJ): A three-times subtracted dispersion relation will be used for the vector form factor. The input phaseshift will be defined using the 
vector form factor in eq.~(\ref{FFs with short distance constraints}) with $\Gamma_{K^\star}(s)$ including only the $K\pi$ cut and resumming also the real part of the loop 
function in the denominator. The unitarized scalar form factor will be used \cite{Jamin:2001zq}. More details can be found in appendix \ref{app}.
\end{itemize}

\section{Predictions for the $\boldsymbol{\tau^-\to K^-\eta\nu_\tau}$ decays}\label{Pred Keta}
We note that eqs.(\ref{FFs with short distance constraints}) also hold for the $\tilde{f}_{+,0}^{K^-\pi}(s)$ form factors (see eq.~(\ref{spectral function}) and comments below, 
as well). Therefore, in principle the knowledge of these form factors in the $K\pi$ system can be transferred to the $K\eta^{(\prime)}$ systems immediately, taking thus 
advantage of the larger statistics accumulated in the former and their sensitivity to the $K^\star(892)$ properties. This is certainly true in the case of the vector form 
factor in its assorted versions and in the scalar Breit-Wigner form factor. However, in the BEJ and JPP scalar form factor one has to bear in mind (see appendix \ref{app:both}) 
that the $KP$ ($P=\pi^0,\,\eta,\,\eta^\prime$) scalar form factors are obtained solving the coupled channel problem which breaks the universality of the $\tilde{f}_0^{K^-P}(s)$ 
form factors as a result of the unitarization procedure. As a consequence, our application of the $\tilde{f}_0^{K^-\eta^{(\prime)}}(s)$ form factors to the 
$\tau^-\to K^-\eta^{(\prime)}\nu_\tau$ decays will provide a test of the unitarized results. Taking into account the explanations in Ref.~\cite{Jamin:2001zq} about the 
difficult convergence of the three-channel problem (mainly because of the smallness of the $K\eta$ contribution and its correlation with the $K\eta^\prime$ channel) this 
verification is by no means trivial, specially regarding the $K\eta^\prime$ channel, where the scalar contribution is expected to dominate the decay width.

In this way, we have predicted the $\tau^-\to K^-\eta\nu_\tau$ branching ratio and differential decay width using the knowledge acquired in the $\tau^-\to (K\pi)^-\nu_\tau$ 
decays. Explicitly:
\begin{itemize}
 \item In the dipole model, we have taken the $K^\star(892)$, $K^\star(1410)$ and $K_0^\star(1430)$ mass and width from the PDG \cite{Beringer:1900zz} -since this compilation 
employs Breit-Wigner parametrizations to determine these parameters- and estimated the relative weight of them using $\gamma=\frac{F_VG_V}{F^2}-1$ (see discussion at the end 
of section \ref{FFs}) \cite{Ecker:1988te}. In this way, we have found $\gamma=-0.021\pm0.031$.
 \item In the JPP parametrization, we have used the best fit results of Ref.~\cite{Jamin:2008qg} for the vector form factor. The scalar form factor has been obtained from the 
solutions (6.10) and (6.11) of Ref.~\cite{Jamin:2001zq} \footnote{The relevant $f_0^{K^-\eta^{(\prime)}}(s)$ unitarized scalar form factors have been coded using tables kindly 
provided by Matthias Jamin.}. The scalar form factors have also been treated alike in the BEJ approach.
 \item In the BEJ representation, one would use the best fit results of Ref.~\cite{Boito:2010me} to obtain our vector form factor. However, we have noticed the strong dependence 
on the actual particle masses of the slope form factor parameters, $\lambda_+^\prime$ and $\lambda_+^{\prime\prime}$. Ref.~\cite{Boito:2010me} used the physical masses in their 
study of $\tau^-\to K_S\pi^-\nu_\tau$ data. On the other hand we focus on the $\tau^-\to K^-P\nu_\tau$ decays. Consequently, the masses should correspond now to $K^-\pi^0$ 
instead of to $K_S\pi^-$. Noteworthy, both the $K^-$ and $\pi^0$ are lighter than the $K_S$ and $\pi^-$ and the corresponding small mass differences, given by isospin breaking, 
are big enough to demand for a corresponding change in the $\lambda_+^{\prime(\prime)}$ parameters. Accepting this, the ideal way to proceed would be to fit the BaBar data on 
$\tau^-\to K^-\pi^0\nu_\tau$ decays \cite{Aubert:2007jh}. Unfortunately, these data are not publicly available yet. For this reason, we have decided to fit Belle data on the 
$\tau^-\to K_S\pi^-\nu_\tau$ decay using the $K^-$ and $\pi^0$ masses throughout. The results can be found in table \ref{Tab:Fake fit}, where they are confronted to the best 
fit results of Ref.~\cite{Boito:2008fq}~\footnote{We display the results of this reference instead of those in Ref.~\cite{Boito:2010me} because we are not using information 
from $K_{\ell3}$ decays in this exercise. Differences are, nonetheless, tiny.}, both of them yield $\chi^2/dof=1.0$ and are given for $s_{cut}=4$ GeV$^2$, although the 
systematic error due to the choice of this energy scale is included in the error estimation. We will use the results in the central column of table \ref{Tab:Fake fit} to give 
our predictions of the $\tau^-\to K^-\eta\nu_\tau$ decays based on the $K\pi$ results.
\begin{table*}[h!]
\begin{center}
\begin{tabular}{|c|c|c|}
\hline
Parameter& Best fit with fake masses & Best fit \cite{Boito:2008fq}\\
\hline
$\lambda_+^\prime\times 10^{3}$&$22.2\pm0.9$& $24.7\pm0.8$\\
$\lambda_+^{\prime\prime}\times 10^{4}$&$10.3\pm0.2$& $12.0\pm0.2$\\
$M_{K^\star}$ (MeV)&$892.1\pm0.6$& $892.0\pm0.9$\\
$\Gamma_{K^\star}$ (MeV)&$46.2\pm0.5$& $46.2\pm0.4$\\
$M_{K^{\star\prime}}$ (GeV)&$1.28\pm0.07$& $1.28\pm0.07$\\
$\Gamma_{K^{\star\prime}}$ (GeV)&$0.16^{+0.10}_{-0.07}$& $0.20^{+0.06}_{-0.09}$\\
$\gamma$&$-0.03\pm0.02$& $-0.04\pm0.02$\\
\hline
\end{tabular}
\caption{\small{Results for the fit to Belle $\tau^-\to K_S\pi^-\nu_\tau$ data \cite{Epifanov:2007rf} with a three-times subtracted dispersion relation including two vector 
resonances in $\widetilde{f}_+^{K\pi}(s)$, according to eq.~(\ref{FFs with short distance constraints}) and resumming the loop function in the denominator (see 
appendix \ref{app:BEJ}), as well as the scalar form factor \cite{Jamin:2001zq}. The middle column is obtained using the masses of the $K^-$ and $\pi^0$} mesons and the last 
column using the $K_S$ and $\pi^-$ masses actually corresponding to the data.}\label{Tab:Fake fit}
\end{center}
\end{table*}
\end{itemize}

\begin{figure}[h!]
\begin{center}
\vspace*{1.25cm}
\includegraphics[scale=0.75]{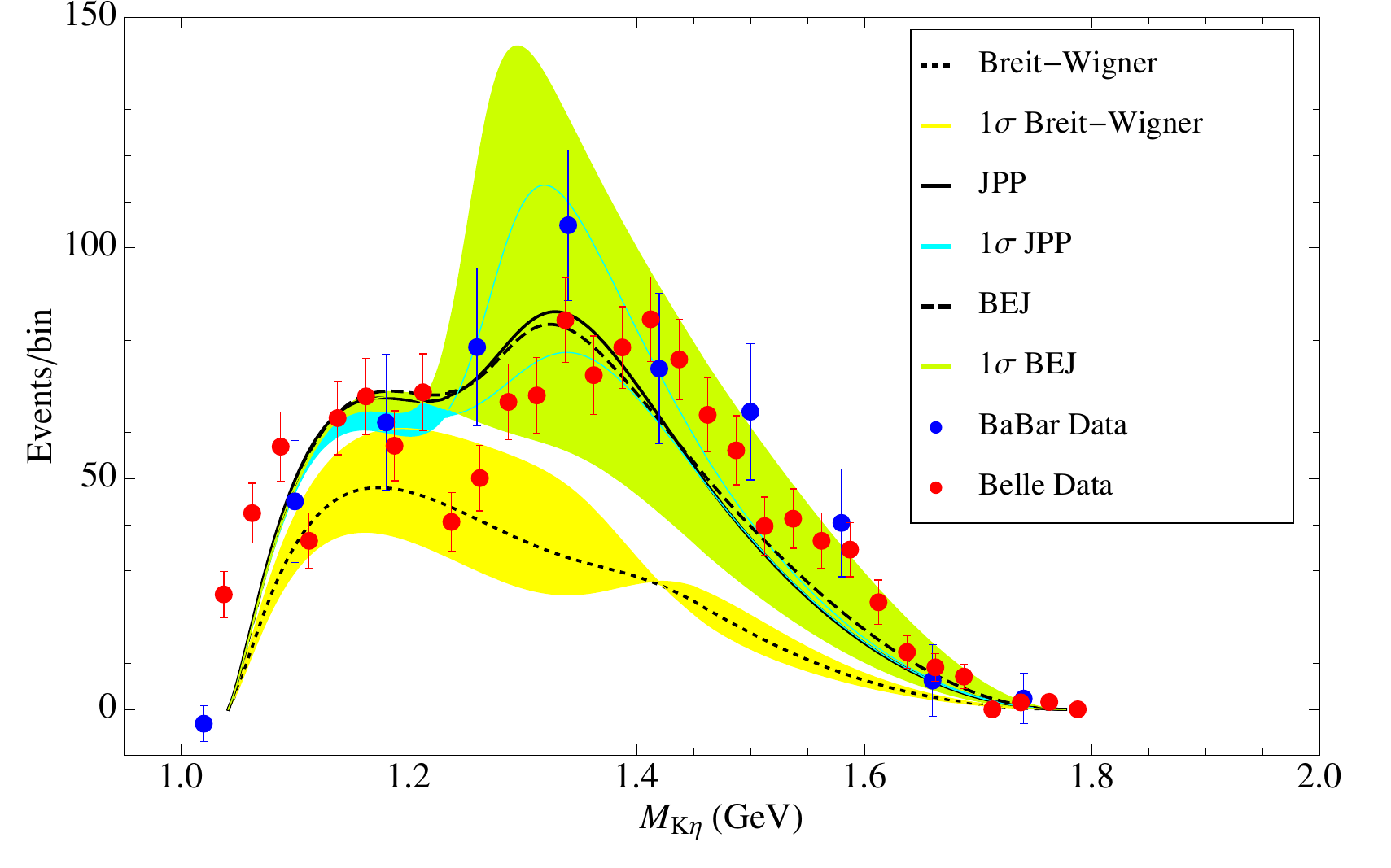}
\caption{\label{fig:Pred_Keta} \small{BaBar (blue) \cite{delAmoSanchez:2010pc} and Belle (red) \cite{Inami:2008ar} data for the $\tau^-\to K^-\eta\nu_\tau$ decays are 
confronted to the predictions obtained in the BW (dotted), JPP (solid) and BEJ (dashed) approaches (see the main text for details) which are shown together with the 
corresponding one-sigma error bands in yellow, light blue and light green, respectively.}}
\end{center}
\end{figure}

Proceeding this way we find the differential decay distributions for the three different approaches considered using eq.~(\ref{spectral function}). 
This one is, in turn, related to the experimental data by using 
\begin{equation}\label{theory_to_experiment}
 \frac{dN_{events}}{dE}\,=\,\frac{d\Gamma}{dE}\frac{N_{events}}{\Gamma_\tau BR(\tau^-\to K^-\eta\nu_\tau)}\Delta E_{bin}\,.
\end{equation}
We thank the Belle Collaboration for providing us with their data \cite{Inami:2008ar}. This was not possible in the case of the BaBar Collaboration \cite{delAmoSanchez:2010pc} 
because the person in charge of the analysis left the field and the data file was lost. We have, however, read the data points from the paper's figures and included this 
effect in the errors. The number of events after background subtraction in each data set are $611$ (BaBar) and $1365$ (Belle) and the corresponding bin widths are $80$ and $25$ 
MeV, respectively. In Fig.\ref{fig:Pred_Keta} we show our predictions based on the $K\pi$ system according to BW, JPP and BEJ. In this figure we have normalized the BaBar data 
to Belle's using eq.~(\ref{theory_to_experiment}). A look at the data shows some tension between both measurements and we notice a couple of strong oscillations of isolated 
Belle data points which do not seem to correspond to any dynamics but rather to an experimental issue or to underestimation of the systematic errors \footnote{We have also 
realized that the first two Belle data points, with non-vanishing entries, are below threshold, a fact which may indicate some problem in the calibration of the hadronic 
system energy or point to underestimation of the background.}. In this plot there are also shown the corresponding one-sigma bands obtained neglecting correlations between 
the resonance parameters and also with respect to other sources of uncertainty, namely $|V_{us}f_+^{K^-\pi^0}(0)|$ and $\theta_P$, whose errors are also accounted for. The 
corresponding branching ratios are displayed in table \ref{Tab:Pred_Keta}, where the $\chi^2/dof$ is also shown. We note that the error correlations corresponding to the fit 
results shown in table \ref{Tab:Fake fit} have been taken into account in BEJ's branching ratio of table \ref{Tab:Pred_Keta}.

It can be seen that the BW model gives a too low decay width and that the function shape is not followed by this prediction, as indicated by the high value of the $\chi^2/dof$ 
that is obtained. On the contrary, the JPP and BEJ predictions yield curves that compare quite well with the data already. Moreover, the corresponding branching fractions are 
in accord with the PDG value within errors. Altogether, this explains the goodness of the $\chi^2/dof$, which is $1.5\leftrightarrow 1.9$. Besides, we notice that the error bands 
are wider in the dispersive representation than in the exponential parametrization, which may be explained by the larger number of parameters entering the former and the more 
complicated correlations between them that were neglected in obtaining Fig.~\ref{fig:Pred_Keta} and the JPP  result in table \ref{Tab:Pred_Keta}.

From these results we conclude that quite likely the BW model is a too rough approach to the problem unless our reference values for $\gamma$ and the $K^\star(1410)$ 
resonance parameters were a bad approximation. We will check this in the next section. On the contrary, the predictions discussed above hint that JPP and BEJ are appropriate 
for the analysis of $\tau^-\to K^-\eta\nu_\tau$ data that we will pursue next.

\begin{table*}[h!]
\begin{center}
\begin{tabular}{|c|c|c|}
\hline
Source& Branching ratio & $\chi^2/dof$\\
\hline
Dipole Model (BW)&$\left(0.78^{+0.17}_{-0.10}\right)\cdot 10^{-4}$& $8.3$\\
JPP&$\left(1.47^{+0.14}_{-0.08}\right)\cdot 10^{-4}$& $1.9$\\
BEJ&$\left(1.49\pm0.05\right)\cdot 10^{-4}$& $1.5$\\
Experimental value&$\left(1.52\pm0.08\right)\cdot 10^{-4}$& -\\
\hline
\end{tabular}
\caption{\small{Predicted branching ratio of the $\tau^-\to K^-\eta\nu_\tau$ decays according to the different approaches used (see the items above eq.~(\ref{theory_to_experiment}) 
for details). The corresponding $\chi^2/dof$ values are also given and the PDG branching fraction is given for reference.}}\label{Tab:Pred_Keta}
\end{center}
\end{table*}

\section{Fits to the $\boldsymbol{\tau^-\to K^-\eta\nu_\tau}$ BaBar and Belle data}\label{Fit Keta}
We have considered different fits to the $\tau^-\to K^-\eta\nu_\tau$ data. In full generality we have assessed that the data is not sensitive either to the low-energy region 
or to the $K^\star(892)$ peak region. This is not surprising, since the threshold for $K^-\eta$ production opens around $1041$ MeV which is some $100$ MeV larger than 
$M_{K^\star}+\Gamma_{K^\star}$, a characteristic energy scale for the $K^\star(892)$ region of dominance. This implies first that the fits are unstable under floating $M_{K^\star}$ 
and $\Gamma_{K^\star}$ (which affects all three approaches) and second that the slopes of the vector form factor, which encode the physics immediately above threshold, can not 
be fitted with $\tau^-\to K^-\eta\nu_\tau$ data (this only concerns BEJ). We have considered consequently fits varying only the $K^\star(1410)$ mass and width and $\gamma$ and 
sticking to the reference values discussed in the previous section for the remaining parameters in every approach.

\begin{figure}[h!]
\begin{center}
\vspace*{1.25cm}
\includegraphics[scale=0.75]{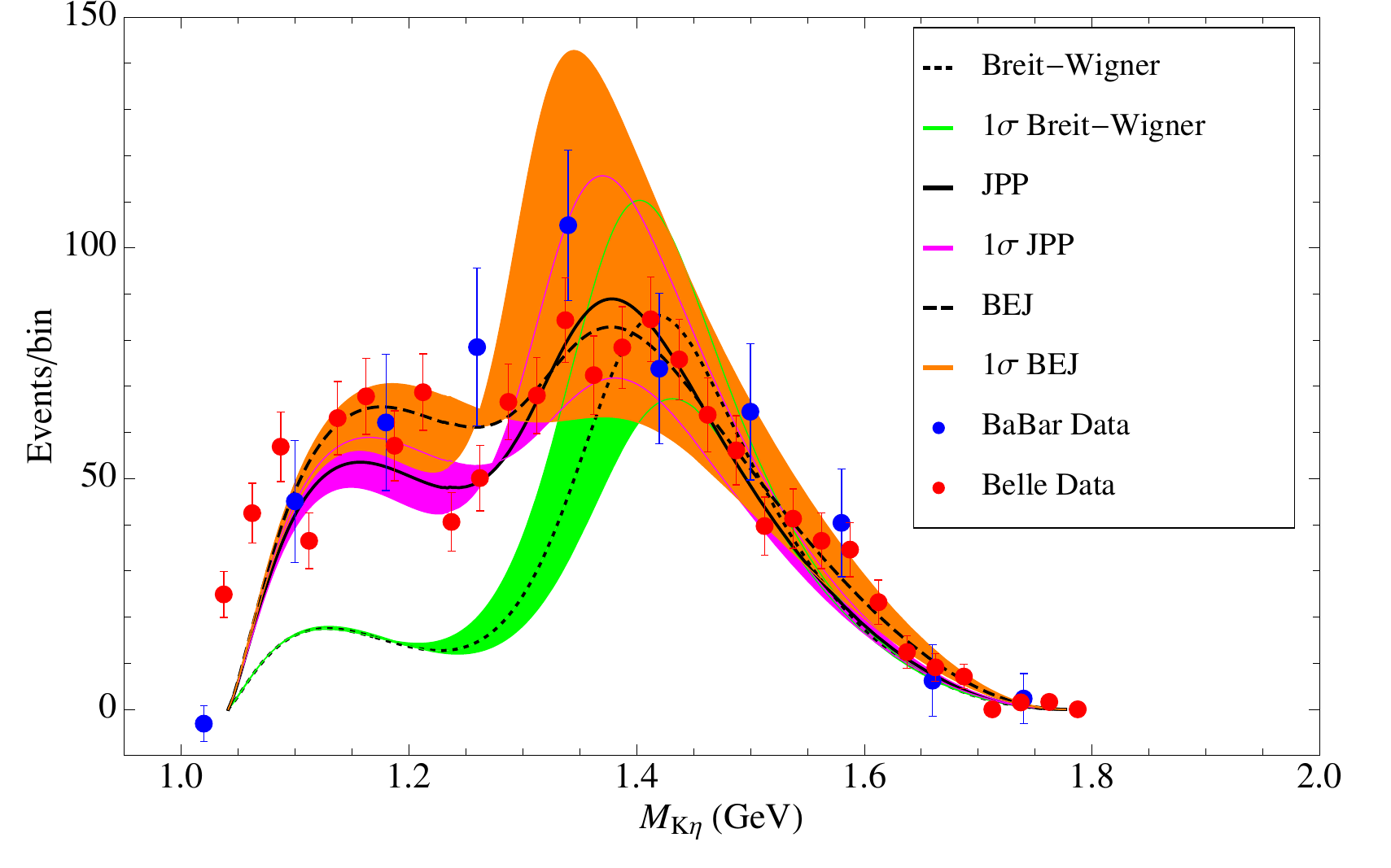}
\caption{\label{fig:Fit_Keta} \small{BaBar (blue) \cite{delAmoSanchez:2010pc} and Belle (red) \cite{Inami:2008ar} data for the $\tau^-\to K^-\eta\nu_\tau$ decays are 
confronted to the best fit results obtained in the BW (dotted), JPP (solid) and BEJ (dashed) approaches (see the main text for details) which are shown together with the 
corresponding one-sigma error bands in light green, pink and orange, respectively. The BW curve corresponds to the one-parameter fit while the JPP and BEJ ones correspond 
to three-parameter fits.}}
\end{center}
\end{figure}

Our best fit results for the branching ratios are written in table \ref{Tab:Fit_Keta}, where the corresponding $\chi^2/dof$ can also be read. These are obtained with the best fit 
parameter values shown in table \ref{Tab:Fit_results}, which can be compared to the reference values, which were used to obtain the predictions in the previous section, that 
are recalled in table \ref{Tab:Fit_Reference}. The corresponding decay distributions with one-sigma error bands attached are plotted in Fig.~\ref{fig:Fit_Keta}.

These results show that the BW model does not really provide a good approximation to the underlying physics for any value of its parameters and should be discarded. Oppositely, 
JPP and BEJ are able to yield quite good fits to the data with values of the $\chi^2/dof$ around one. This suggests that the simplified treatment of final state interactions 
in BW, which misses the real part of the two-meson rescatterings and violates analyticity by construction, is responsible for the failure.

A closer look to the fit results using JPP and BEJ in tables \ref{Tab:Fit_Keta} and \ref{Tab:Fit_results} shows that:
\begin{itemize}
 \item Fitting $\gamma$ alone is able to improve the quality of both approaches by $15\leftrightarrow 20\%$. The fitted values are consistent with the reference ones (see table 
\ref{Tab:Fit_Reference}): in the case of BEJ at one sigma, being the differences in JPP slightly larger than that only. This is satisfactory because both the 
$\tau^-\to (K\pi)^-\nu_\tau$ and the $\tau^-\to K^-\eta\nu_\tau$ decays are sensitive to the interplay between the first two vector resonances and contradictory results would 
have casted some doubts on autoconsistency.
 \item When the $K^\star(1410)$ parameters are also fitted the results improve by $\sim13\%$ in JPP and by $\sim33\%$ in BEJ. This represents a reduction of the $\chi^2/dof$ 
by $\sim26\%$ in JPP and by $\sim50\%$ in BEJ. It should be noted that the three-parameter fits do not yield to physical results in BW. Specifically, $K^\star(1410)$ mass and 
width tend to the $K^\star(892)$ values and $|\gamma|$ happens to be one order of magnitude larger than the determinations in the literature. Therefore we discard this 
result. We also notice that although the branching ratios of both JPP and BEJ (which have been obtained taking into account the parameter fit correlations) are in agreement 
with the PDG value, the JPP branching ratios tend to be closer to its lower limit while BEJ is nearer to the upper one. It can be observed that the deviations of the 
three-parameter best fit values with respect to the default ones lie within errors in BEJ, as it so happens with $\Gamma_{K^{\star\prime}}$ in JPP. However, there are small 
tensions between the reference and best fit values of $M_{K^{\star\prime}}$ and $\gamma$ in JPP.
\end{itemize}

These results are plotted in Fig.~\ref{fig:Fit_Keta}. Although the BW curve has improved with respect to Fig.~\ref{fig:Pred_Keta} and seems to agree well with the data in the 
higher-energy half of the spectrum, it fails completely at lower energies. On the contrary, JPP and BEJ provide good quality fits to data which are satisfactory along the 
whole phase space. We note that JPP goes slightly below BEJ and its error band is again narrower possibly due to having less parameters. BEJ errors include the systematics 
associated to changes in $s_{cut}$ which is slightly enhanced with respect to the $K\pi$ case.

Despite the vector form factor giving the dominant contribution to the decay width, the scalar form factor is not negligible and gives $\sim(3\leftrightarrow4)\%$ 
of the branching fraction in the JPP and BEJ cases. In the BW model this contribution is $\sim7\%$.

\begin{table*}[h!]
\begin{center}
\begin{tabular}{|c|c|c|}
\hline 
Source & Branching ratio & $\chi^2/dof$\\
\hline
Dipole Model (BW) (Fit $\gamma$)&$\left(0.96^{+0.21}_{-0.15}\right)\cdot10^{-4}$& $5.0$\\
Dipole Model (BW) (Fit $\gamma$, $M_{K^{\star\prime}}$, $\Gamma_{K^{\star\prime}})$ &Unphysical result& -\\
JPP (Fit $\gamma$)&$\left(1.50^{+0.19}_{-0.11}\right)\cdot 10^{-4}$& $1.6$\\
JPP (Fit $\gamma$, $M_{K^{\star\prime}}$, $\Gamma_{K^{\star\prime}})$&$\left(1.42\pm0.04\right)\cdot 10^{-4}$& $1.4$\\
BEJ (Fit $\gamma$)&$\left(1.59^{+0.22}_{-0.16}\right)\cdot 10^{-4}$& $1.2$\\
BEJ (Fit $\gamma$, $M_{K^{\star\prime}}$, $\Gamma_{K^{\star\prime}})$ &$\left(1.55\pm0.08\right)\cdot 10^{-4}$& $0.8$\\
Experimental value&$\left(1.52\pm0.08\right)\cdot 10^{-4}$& -\\
\hline
\end{tabular}
\caption{\label{Tab:Fit_Keta} \small{The branching ratios and $\chi^2/dof$ obtained in BW, JPP and BEJ fitting $\gamma$ only and also the $K^\star(1410)$ parameters are 
displayed. Other parameters were fixed to the reference values used in section \ref{Pred Keta}. The PDG branching fraction is also given for reference.}}
\end{center}
\end{table*}

\begin{table}
\begin{center}
\begin{tabular}{|c|c|c|c|c|c|c|c}
\hline
\backslashbox{Fitted value}{Approach}&Dipole Model (BW)&JPP&BEJ\cr
\hline
$\gamma$&$-0.174\pm0.007$&$-0.063\pm0.007$&$-0.041\pm0.021$\cr
\hline
$\gamma$&Unphysical&$-0.078^{+0.012}_{-0.014}$&$-0.051^{+0.012}_{-0.036}$\cr
$M_{K^{\star'}}$ (MeV) &best fit&$1356\pm11$&$1327^{+30}_{-38}$\cr
$\Gamma_{K^{\star'}}$ (MeV) &parameters&$232^{+30}_{-28}$&$213^{+72}_{-118}$\cr
\hline
\end{tabular}
\caption{\label{Tab:Fit_results} \small{The best fit parameter values corresponding to the different alternatives considered in table \ref{Tab:Fit_Keta} are given. These can 
be compared to the reference values, which are given in table \ref{Tab:Fit_Reference}. BEJ results for the mass and width of the $K^{\star}(1410)$ correspond to pole values, 
while JPP figures are given for the model parameter as in the original literature.}}
\end{center}
\end{table}

\begin{table}
\begin{center}
\begin{tabular}{|c|c|c|c|c|c|c|c}
\hline
\backslashbox{Reference value}{Approach}&Dipole Model (BW)&JPP&BEJ\cr
\hline
$\gamma$&$-0.021\pm0.031$&$-0.043\pm0.010$&$-0.029\pm0.017$\cr
$M_{K^{\star'}}$ (MeV) & $1414\pm15$ &$1307\pm17$&$1283\pm65$\cr
$\Gamma_{K^{\star'}}$ (MeV) & $232\pm21$ &$206\pm49$&$163\pm68$\cr
\hline
\end{tabular}
\caption{\label{Tab:Fit_Reference} \small{Reference values (used in section \ref{Pred Keta}) corresponding to the best fit parameters appearing in table \ref{Tab:Fit_results}. 
Again BEJ results are pole values and JPP ones are model parameters. The latter are converted to resonance pole values in section \ref{Concl}, where the determination of the 
$K^{\star}(1410)$ pole parameters is given.}}
\end{center}
\end{table}

The JPP model values appearing in tables \ref{Tab:Fit_results} and \ref{Tab:Fit_Reference} can be translated to pole values along the lines discussed in 
Ref.~\cite{Escribano:2002iv}. This yields $M_{K^{\star\prime}}=1332^{+16}_{-18}\,,\,\Gamma_{K^{\star\prime}}=220^{+26}_{-24}$ for the best fit values and 
$M_{K^{\star\prime}}=1286^{+26}_{-28}\,,\,\Gamma_{K^{\star\prime}}=197^{+41}_{-45}$ for the reference values, where all quantities are given in MeV. Remarkable 
agreement is found between our best fit values in the JPP and BEJ cases, since the latter yields $M_{K^{\star\prime}}=1327^{+30}_{-38}\,,
\,\Gamma_{K^{\star\prime}}=213^{+72}_{-118}$. From the detailed study of the $\pi\pi$, $K\pi$ (in the quoted literature) and $K\eta$ systems (in this paper) within JPP and BEJ, 
one can conclude generally that the dispersive form factors allow a better description of the data while the exponential parametrizations lead to the determination of the 
resonance pole values with smaller errors. Both things seem to be due to the inclusion of the subtraction constants as extra parameters in the fits within the dispersive 
representations.

\section{Predictions for the $\boldsymbol{\tau^-\to K^-\eta^\prime\nu_\tau}$ decays}\label{Pred Ketap}
\begin{figure}[h!]
\begin{center}
\vspace*{1.25cm}
\includegraphics[scale=0.75]{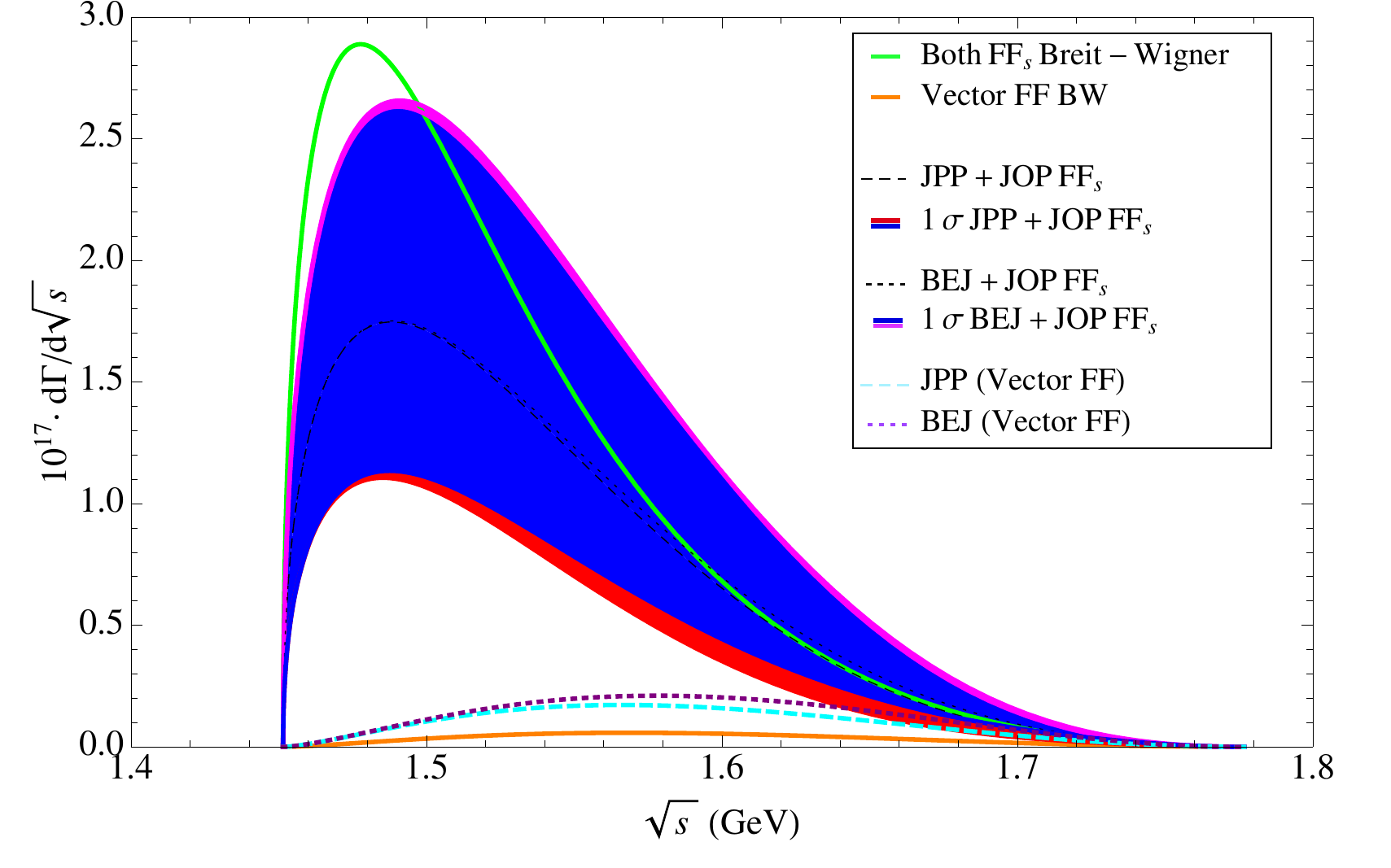}
\caption[]{\label{fig:Ketap} \small{The predicted $\tau^-\to K^-\eta^\prime\nu_\tau$ decay width according to BW (green, its big uncertainty is not shown for clarity of the 
figure), JPP (blue with lower band in red) and BEJ (blue with upper part in pink) is shown. In these last two the scalar form factor corresponds to Ref.~\cite{Jamin:2006tj}, 
which is represented by the author's initials, JOP, in the figure's legend. The corresponding vector form factor contributions, which are subleading are plotted 
in orange (solid), blue (dashed) and purple (dotted).}}
\end{center}
\end{figure}

We can finally profit from our satisfactory description of the $\tau^-\to K^-\eta\nu_\tau$ decays and predict the $\tau^-\to K^-\eta^\prime\nu_\tau$ decay observables, where 
there is only the upper limit fixed at ninety percent confidence level by the BaBar Collaboration \cite{Lees:2012ks}, $BR<4.2\cdot10^{-6}$. We have done this for our best fit 
results in the BW (one-parameter fit) JPP and BEJ (three-parameter fits) cases. The corresponding results are plotted in Fig.~\ref{fig:Ketap} and the branching ratios can be 
read from table \ref{Tab:Pred_Ketap}. In the figure we can see that the decay width is indeed dominated by the scalar contribution \footnote{In principle, both the scalar and 
vector $K\eta^\prime$ form factors are suppressed since they are proportional to $\sin\theta_P$. However, the unitarization procedure of the scalar form factor enhances it 
sizeably \cite{Jamin:2001zq} due to the effect of the coupled inelastic channels.} \footnote{The suppression of the vector contribution makes that the predicted values using 
information from the $K\pi$ system and the one-parameter fits with JPP and BEJ are very similar to the results in table \ref{Tab:Pred_Ketap}. For this reason we do not show them.}. 
In fact, the vector form factor contributes in the range $(9\leftrightarrow15)\%$ to the corresponding branching ratio. Although we keep the BW prediction for reference, we do 
not draw the associated (large) error band for the sake of clarity in the figure taking into account its wrong description of the $K\eta$ system shown in the previous 
section. As the scalar form factor dominates the decay width and we are using the same one in JPP and BEJ, the differences between them are tiny (and the errors, of order one 
third, are the same in table \ref{Tab:Pred_Ketap}). As expected from the results in the $\tau^-\to K^-\eta\nu_\tau$ decays, BEJ gives the upper part of the error band while 
JPP provides the lower one. We are looking forward to the discovery of this decay mode to verify our predictions. A priori one may forecast some departure from it because of 
the effect of the poorly known elastic and $K\eta$ channels in meson-meson scattering, which affects the solution of the coupled system of integral equations and specially 
the value of the $K^-\eta^\prime$ scalar form factor, that is anyway suppressed to some extent.

\begin{table*}[h!]
\begin{center}
\begin{tabular}{|c|c|}
\hline 
Source & Branching ratio\\
\hline
Dipole Model (BW) (Fit)&$(1.45^{+3.80}_{-0.87})\cdot10^{-6}$\\
JPP (Fit)&$(1.00^{+0.37}_{-0.29})\cdot10^{-6}$\\
BEJ (Fit)&$(1.03^{+0.37}_{-0.29})\cdot10^{-6}$\\
Experimental bound&<$4.2\cdot10^{-6}$ at $90\%$ C.L.\\
\hline
\end{tabular}
\caption{\label{Tab:Pred_Ketap} \small{Predicted branching ratios for the $\tau^-\to K^-\eta^\prime\nu_\tau$ decays. The BaBar upper limit is also shown \cite{Lees:2012ks}.}}
\end{center}
\end{table*}

\begin{figure}[!h]
\begin{center}
\vspace*{1.25cm}
\includegraphics[scale=1.25]{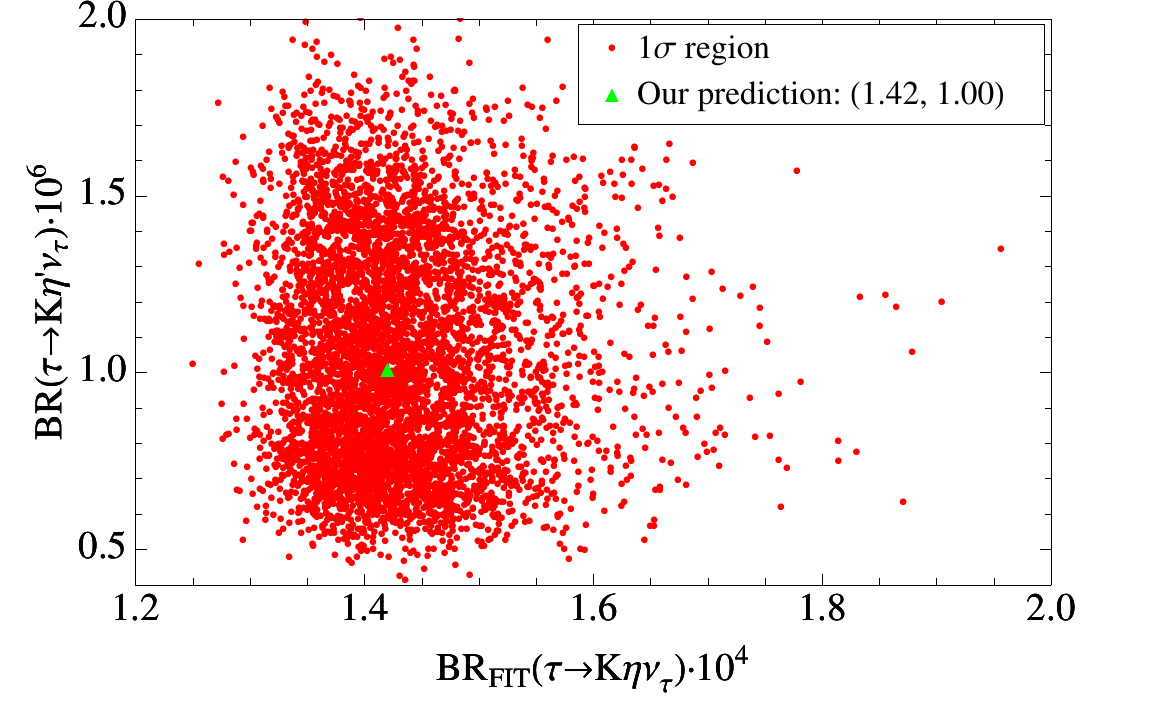}
\caption[]{\label{fig:Correlation} \small{The correlation between the $\tau^-\to K^-\eta\nu_\tau$ and $\tau^-\to K^-\eta^\prime\nu_\tau$ branching ratios is plotted according 
to the best fit JPP result at one sigma. Correlations between the parameters are neglected. According to expectations, no sizable correlation between both decay modes is 
observed.}}
\end{center}
\end{figure}

In Fig.~\ref{fig:Correlation} we also plot the correlation between the $\tau^-\to K^-\eta\nu_\tau$ and $\tau^-\to K^-\eta^\prime\nu_\tau$ branching ratios according to the best fit JPP 
result at one sigma. The correlations between the parameters are neglected. Since the vector (scalar) form factor dominates the former (latter) decays and their parameters are 
independent the plot does not show any sizeable correlation between both measurements, as expected. As a result, if new data on the $\tau^-\to K^-\eta^\prime\nu_\tau$ decays 
demand a more careful determination of the $f_0^{K^-\eta^\prime}(s)$ unitarized form factor this will leave almost unaffected the results obtained for the $\tau^-\to K^-\eta\nu_\tau$ 
channel.

\section{Conclusions}\label{Concl}
Hadronic tau decays are an ideal scenario to learn about the non-perturbative character of the strong interactions in rather clean conditions. In this work, we have studied 
the $\tau^-\to K^-\eta^{(\prime)}\nu_\tau$ decays motivated by the recent measurements performed by the BaBar \cite{delAmoSanchez:2010pc, Lees:2012ks} and Belle Collaborations 
\cite{Inami:2008ar}. These decays allow the application of the knowledge acquired in the study of $\tau^-\to (K\pi)^-\nu_\tau$ decays. In particular, the $K\eta$ decay is 
sensitive to the parameters of the $K^\star(1410)$ resonance and to its interplay with the $K^\star(892)$ meson, while the $K\eta^\prime$ decay is an appropriate place to 
test the unitarization of the strangeness-changing scalar form factors in three coupled-channel case.

We have defined with detail the (tilded) scalar and vector form factors and we have gone through the steps of their calculation within Chiral Perturbation Theory including 
the lightest resonances as explicit degrees of freedom and showed that the results are written in a more compact way using the tilded form factors. Then we have discussed 
different options according to the treatment of final-state interactions. Specifically, there is the dipole Breit-Wigner (BW) model, which neglects the real part of the two-meson 
loop function violating analyticity at next-to-leading order; there is the exponential parametrization (JPP) where this real part of the loop is resummed through an Omn\`es 
exponentiation, which violates analyticity at the next order; and there is the dispersive representation (BEJ), which resums the whole loop function in the denominators, 
where analyticity holds exactly.

In our case, an additional difficulty is that the elastic approach is not valid in any region of the phasespace, since the $K\pi$ channel is 
open well below the $K\eta^{(\prime)}$ channels. In JPP this is not an issue, since one simply adds the corresponding contribution of these channels to the width and real part 
of the loop function. However, in BEJ it prevents an approach which does not include inelasticities and the effect of coupled channels. Being conscious of this, we have 
nevertheless attempted a dispersive representation of the $K\eta^{(\prime)}$ vector form factors were the input phaseshift is obtained using the elastic approximation and, to 
our surprise, it has done an excellent job in its confrontation to the $K\eta$ data. In the light of more accurate measurements it may become necessary to improve this 
treatment in the future. Very good agreement has also been found using JPP but BW has failed in this comparison. In the JPP and BEJ fits to the $K\eta$ channel the scalar 
form factor was obtained solving dispersion relations for the three-body problem.

We have checked that the $K\eta^{(\prime)}$ channels are not sensitive either to the $K^\star(892)$ parameters or to the slopes of the form factor, $\lambda_+^{\prime(\prime)}$ 
(BEJ). We have borrowed this information from the $K\pi$ system. This task was straightforward in BW and JPP although in BEJ we noticed that the $\lambda_+^{\prime(\prime)}$ 
parameters were sensitive to isospin breaking effects that we had to account for. Once this was done we could fit the $K^\star(1410)$ resonance pole parameters and its relative 
weight with respect to the $K^\star(892)$ meson, $\gamma$. Our results for these, with masses and widths in MeV, are
\begin{equation}
 M_{K^{\star\prime}}\,=\,1327^{+30}_{-38},\quad\Gamma_{K^{\star\prime}}\,=\,213^{+72}_{-118},\quad \gamma\,=\,-0.051^{+0.012}_{-0.036}\,,
\end{equation}
in the dispersive representation (BEJ) and 
\begin{equation}
 M_{K^{\star\prime}}\,=\,1332^{+16}_{-18},\quad\Gamma_{K^{\star\prime}}\,=\,220^{+26}_{-24},\quad \gamma\,=\,-0.078^{+0.012}_{-0.014}\,,
\end{equation}
for the exponential parametrization (JPP). Our determination of these parameters has shown to be competitive with its extraction from the $\tau^-\to (K\pi)^-\nu_\tau$ decays. 
To illustrate this point, we average the JPP and BEJ determinations from the $K\pi$ \cite{Jamin:2008qg, Boito:2010me} and $K\eta$ systems, respectively, to find
\begin{equation}
 M_{K^{\star\prime}}\,=\,1277^{+35}_{-41},\quad\Gamma_{K^{\star\prime}}\,=\,218^{+95}_{-66},\quad \gamma\,=\,-0.049^{+0.019}_{-0.016}\,,
\end{equation}
from $K\pi$
and 
\begin{equation}
 M_{K^{\star\prime}}\,=\,1330^{+27}_{-41},\quad\Gamma_{K^{\star\prime}}\,=\,217^{+68}_{-122},\quad \gamma\,=\,-0.065^{+0.025}_{-0.050}\,,
\end{equation}
from $K\eta$. We have thus opened an alternative way of determining these parameters. New, more precise data on the $\tau^-\to (K\pi)^-\nu_\tau$ and 
$\tau^-\to K^-\eta\nu_\tau$ decays will make possible a more accurate determination of these parameters.

Finally we have benefited from this study of the $\tau^-\to K^-\eta\nu_\tau$ decays and applied it to the $\tau^-\to K^-\eta^{\prime}\nu_\tau$ decays, were our predictions 
respect the upper limit found by BaBar and hint to the possible discovery of this decay mode in the near future.

In this way we consider that we are in position of providing TAUOLA with theory-based currents that can describe well the $\tau^-\to K^-\eta^{(\prime)}\nu_\tau$ decays, 
based on the exponential parametrization developed by JPP and the dispersive representation constructed by BEJ.

To conclude, differential distributions of hadronic tau decays provide important information for testing diverse form factors and extracting the corresponding parameters 
increasing our knowledge of hadronization in the low-energy non-perturbative regime of QCD. It will be interesting to see if our predictions for the $\tau^-\to K^-\eta^{\prime}\nu_\tau$ 
decays are corroborated and if more precise data on the $\tau^-\to K^-\eta\nu_\tau$ decays demand a more refined treatment. Finally, we emphasize the need of giving pole 
resonance parameters irrespective of the approach employed, either in a theorists' article or in a publication by an experimental collaboration.

\appendix
\section{Form factors for the JPP and BEJ approaches}\label{app}
We refer the reader to the detailed discussions on the subject that are given in Refs.~\cite{Guerrero:1997ku, Jamin:2006tk, Jamin:2008qg} (\ref{app:JPP}), 
\cite{Pich:2001pj, Boito:2008fq, Boito:2010me, Dumm:2013zh} (\ref{app:BEJ}) and \cite{Jamin:2000wn, Jamin:2001zq, Jamin:2001zr, Jamin:2006tj} (\ref{app:both}). Here we 
only give the minimum material that is needed to understand the different approaches that have been employed in our analysis in sections \ref{Pred Keta}-\ref{Pred Ketap}.
\subsection{JPP vector form factor}\label{app:JPP}
The exponential parametrization was developed in the famous Guerrero-Pich paper \cite{Guerrero:1997ku} devoted to the pion vector form factor at the end of last century. 
We will adopt here the discussion to the $K\pi$ case which determines the $K\eta^{(\prime)}$ processes.

The key point in obtaining the Omn\`es solution is that in the elastic region Watson final-state theorem relates the imaginary part of the vector form factor to the 
partial wave amplitude for $K\pi$ scattering with spin one and isospin one-half, $T^{1/2}_1(s)$. In fact, in this region both phases are equal, which allows to write an 
n-subtracted dispersion relation which has the well-known Omn\`es solution
\begin{equation}
\label{omel}
f_+^{K\pi}(s) \,=\, P_n(s) \exp \Biggl\{ \frac{s^n}{\pi}\!
\int\limits^\infty_{s_{\rm thr}}\!\!ds'\, \frac{\delta_1^{1/2}(s')}
{(s')^n(s'-s-i\epsilon)}\Biggr\} \ ,
\end{equation}
where
\begin{equation}
\label{Pns}
\log P_n(s) \,=\, \sum\limits_{k=0}^{n-1} \,\alpha_k \,
\frac{s^k}{k!}
\end{equation}
is the corresponding subtraction polynomial. The subtraction constants $\alpha_k$ are given by~\footnote{More general formulae with subtractions at an arbitrary point $s=s_0$ 
can for example be found in Ref.~\cite{Pallante:2000hk}.}
\begin{equation}
\alpha_k \, =\, \frac{d^k}{ds^k}\log f_+^{K\pi}(s)\biggr|_{s=0} \ .
\end{equation}
Using the leading-order $\chi PT$ result in the integral (\ref{omel}) generates the $\chi PT$ one-loop function at the next order. In this way, the Omn\`es formula 
provides an exponentiation of the chiral logarithmic corrections. The ambiguity in the non-logarithmic part of the Omn\`es relation can be resolved to a large extent 
by matching it to the $R\chi T$ result yielding
\begin{equation}
 f_+^{K\pi}(s)\,=\,\frac{M_{K^\star}^2}{M_{K^\star}^2-s}\mathrm{exp}\left\lbrace\frac{3}{2}\left[\widetilde{H}_{K\pi}(s)+\widetilde{H}_{K\eta}(s)\right]\right\rbrace\,,
\end{equation}
where $\widetilde{H}_{PQ}(s)$ subtracts the contribution of the local term at next-to-leading order in $\chi PT$ from the untilded function \footnote{$H_{PQ}(s)$ is the 
standard Gasser and Leutwyler's two-particle loop function \cite{Gasser:1983yg}.} to avoid double counting, since this term is recovered upon integration of the vector 
resonances in the chosen formalism.

The problem, however, comes when the resonance width is included (as it should to avoid the divergent behaviour of the denominator at the resonance mass). In Ref.~\cite{Guerrero:1997ku} 
the imaginary part of the loop function (giving the resonance width) was shifted to the denominator by hand, which resulted in an expression analogous to
\begin{equation}
 f_+^{K\pi}(s)\,=\,\frac{M_{K^\star}^2}{M_{K^\star}^2-s-iM_{K^\star}\Gamma_{K^\star}(s)}\mathrm{exp}\left\lbrace\frac{3}{2}Re\left[\widetilde{H}_{K\pi}(s)+\widetilde{H}_{K\eta}(s)\right]\right\rbrace\,.
\end{equation}
This approach was also followed in the $K\pi$ analyses. In this way, analyticity holds perturbatively up to next-to-leading order.
\subsection{BEJ vector form factor}\label{app:BEJ}
Analyticity warrants that the vector form factor must satisfy a dispersion relation and unitarity that the dispersion relation admits a well-known closed-form solution 
within the elastic approximation referred as the Omn\`es representation. This simple and elegant solution is unrealistic at the practical level since (as a consequence of 
analyticity) it demands the detailed knowledge of the form factor phase up to infinity. This problem is circumvented by considering additional subtractions (one -the normalization 
at the origin- is needed for the convergence of the form factor and is best determined from lattice QCD) which increase the weight of the lower-energy region and damp the 
problematic higher-energy zone, since an n-times-subtracted form factor exhibits a suppression of $s^{-(n+1)}$ in the integrand. This results in a transfer of the information 
that was previously encoded in the high-energy part of the integral into $n-1$ subtraction constants.
The analyses of the $\pi\pi$ \cite{Pich:2001pj, Dumm:2013zh} and $K\pi$ \cite{Boito:2008fq, Boito:2010me} vector form factors within this framework shows an optimal 
description of the data with three subtractions. This result will be followed using
\begin{equation}
 \widetilde{f}_+(s)\,=\,\mathrm{exp}\left[\alpha_1\frac{s}{m_\pi^2}+\frac{1}{2}\alpha_2\frac{s^2}{m_\pi^4}+\frac{s^3}{\pi}\int_{s_{K\pi}}^{s_{cut}}ds^\prime
\frac{\delta(s^\prime)}{(s^\prime)^3(s^\prime-s-i0)}\right]\,,
\end{equation}
where $s_{K\pi}=(m_K+m_\pi)^2$ \footnote{The values of the masses that are actually used in this relation are discussed in section \ref{Pred Keta}.} and the two subtraction 
constants are related to the low-energy expansion of the $\widetilde{f}_+(s)$ form factor:
\begin{equation}
 \widetilde{f}_+(s)\,=\,1+\lambda_+^\prime\frac{s}{m_\pi^2}+\frac{1}{2}\lambda_+^{\prime\prime}\frac{s^2}{m_\pi^4}+...\,,
\end{equation}
while the value of the cut-off, $s_{cut}$, should in principle be varied to estimate the associated systematic error.

The input phase, $\delta(s)$, is obtained as
\begin{equation}
 \delta(s)\,=\,\mathrm{tan}^{-1}\left[\frac{\mathrm{Im}\widetilde{f}_+(s)}{\mathrm{Re}\widetilde{f}_+(s)}\right]\,,
\end{equation}
where $\widetilde{f}_+(s)$ resums the real part of the two-point loop function in the denominator \cite{Jamin:2000wn, Oller:2000ma}:
\begin{equation}\label{Tilded VFF BEJ}
 \widetilde{f}_+(s)\,=\,\frac{m_{K^\star}^2-\kappa_{K^\star}\widetilde{H}_{K\pi}(0)+\gamma s}{D(m_{K^\star},\,\gamma_{K^\star})}-\frac{\gamma s}{D(m_{K^{\star\prime}},\,
\gamma_{K^{\star^\prime}})}\,.
\end{equation}
The denominators in eq.~(\ref{Tilded VFF BEJ}) are
\begin{equation}
 D(m_n,\gamma_n)\equiv m_n^2-s-\kappa_n Re\left[H_{K\pi}(s)\right]-im_n\gamma_n(s)\,,
\end{equation}
where
\begin{equation}
 \kappa_n\,=\,\frac{192\pi F_KF_\pi}{\sigma^3(m_n^2)}\frac{\gamma_n}{m_n}\,,\quad \gamma_n(s)\,=\,\gamma_n\frac{s}{m_n^2}\frac{\sigma^3_{K\pi}(s)}{\sigma^3_{K\pi}(m_n^2)}\,,
\end{equation}
and $\sigma(m_P^2)=\sigma_{PP}(s)=\sqrt{1-\frac{4m_P^2}{s}}$ is the two-body phase-space factor.
\subsection{Scalar form factor in both approaches}\label{app:both}
In Ref.\cite{Jamin:2001zq} the multi-channel Muskelishivili-Omn\`es problem for three channels ($K\pi$, $K\eta$, $K\eta^\prime$ for $i=1,2,3$) is solved. Each of the scalar 
form factors $f_0^i(s)$ is then coupled to the others via
\begin{equation}\label{OM}
 f_0^i(s)\,=\,\frac{1}{\pi}\sum_{j=1}^3\int_{s_i}^\infty ds^\prime\frac{\sigma_j(s^\prime)f_0^j(s^\prime)t_0^{i\to j}(s^\prime)^\star}{(s^\prime-s-i0)}\,,
\end{equation}
where $s_i$ is the threshold for channel $i$ and $t_0^{i\to j}$ are partial wave $T$-matrix elements for the $i\to j$ scattering. The unitarized form factors are obtained 
solving the coupled dispersion relations arising from eq.~(\ref{OM}) imposing chiral symmetry constraints and using $T$-matrix elements from Ref.\cite{Jamin:2000wn} providing 
an accurate description of meson-meson scattering data. In the elastic approximation, eq.~(\ref{OM}) reduces to the usual single-channel Omn\`es equation.
\acknowledgments
We thank very much M.~Jamin and J.~Portol\'es for their careful critical reading of our draft. Discussions with A.~Pich and R.~Sobie on this topic are very much appreciated. 
We are grateful to the Belle Collaboration for providing us with their data and, in particular, to Kenji Inami for correspondence concerning our analysis. This work was 
supported in part by the FPI scholarship BES-2012-055371 (S.G-S), the Ministerio de Ciencia e Innovaci\'on under grants FPA2011-25948 and AIC-D-2011-0818, the European 
Commission under the 7th Framework Programme through the ``Research Infrastructures'' action of the ``Capacities'' Programme Call: FP7-INFRA-STRUCTURES-2008-1 (Grant 
Agreement N. 227431), the Spanish Consolider-Ingenio 2010 Programme CPAN (CSD2007-00042), and the Generalitat de Catalunya under grant SGR2009-00894.


\begin{thebibliography}{99}
\bibitem{Braaten:1991qm}
  E.~Braaten, S.~Narison and A.~Pich,
  Nucl.\ Phys.\ B {\bf 373} (1992) 581.

\bibitem{Braaten:1988hc}
  E.~Braaten,
  Phys.\ Rev.\ Lett.\  {\bf 60} (1988) 1606.

\bibitem{Braaten:1988ea}
  E.~Braaten,
  Phys.\ Rev.\ D {\bf 39} (1989) 1458.

\bibitem{Braaten:1990ef}
  E.~Braaten and C.~-S.~Li,
  Phys.\ Rev.\ D {\bf 42} (1990) 3888.

\bibitem{Narison:1988ni}
  S.~Narison and A.~Pich,
  Phys.\ Lett.\ B {\bf 211} (1988) 183.

\bibitem{Pich:1989pq}
  A.~Pich,
  Conf.\ Proc.\ C {\bf 890523} (1989) 416.

\bibitem{Davier:2005xq}
  M.~Davier, A.~Hocker and Z.~Zhang,
  Rev.\ Mod.\ Phys.\  {\bf 78} (2006) 1043.

\bibitem{Pich:2013??}
A. Pich, 
``Precision Tau Physics'',
to be published as a Review in ``Progress in Particle and Nuclear Physics''.

\bibitem{Davier:2008sk}
  M.~Davier, S.~Descotes-Genon, A.~Hocker, B.~Malaescu and Z.~Zhang,
  Eur.\ Phys.\ J.\ C {\bf 56} (2008) 305.

\bibitem{Beneke:2008ad}
  M.~Beneke and M.~Jamin,
  JHEP {\bf 0809} (2008) 044.

\bibitem{Pich:2011bb}
  A.~Pich,
 arXiv:1107.1123 [hep-ph]. Published in the Proc. of the Workshop on Precision Measurements of $\alpha_S$
 9-11 Feb 2011. Munich, Germany.

\bibitem{Boito:2012cr}
  D.~Boito, M.~Golterman, M.~Jamin, A.~Mahdavi, K.~Maltman, J.~Osborne and S.~Peris,
  Phys.\ Rev.\ D {\bf 85} (2012) 093015.

\bibitem{Barate:1999hj}
  R.~Barate {\it et al.}  [ALEPH Collaboration],
  Eur.\ Phys.\ J.\ C {\bf 11} (1999) 599.

\bibitem{Abbiendi:2004xa}
  G.~Abbiendi {\it et al.}  [OPAL Collaboration],
  Eur.\ Phys.\ J.\ C {\bf 35} (2004) 437.

\bibitem{Maltman:2008ib}
  K.~Maltman, C.~E.~Wolfe, S.~Banerjee, J.~M.~Roney and I.~Nugent,
  Int.\ J.\ Mod.\ Phys.\ A {\bf 23} (2008) 3191.

\bibitem{Antonelli:2013usa}
  M.~Antonelli, V.~Cirigliano, A.~Lusiani and E.~Passemar,
  arXiv:1304.8134 [hep-ph].

\bibitem{Chetyrkin:1998ej}
  K.~G.~Chetyrkin, J.~H.~Kuhn and A.~A.~Pivovarov,
  Nucl.\ Phys.\ B {\bf 533} (1998) 473

\bibitem{Pich:1999hc}
  A.~Pich and J.~Prades,
  JHEP {\bf 9910} (1999) 004.

\bibitem{Korner:2000wd}
  J.~G.~Korner, F.~Krajewski and A.~A.~Pivovarov,
  Eur.\ Phys.\ J.\ C {\bf 20} (2001) 259.

\bibitem{Kambor:2000dj}
  J.~Kambor and K.~Maltman,
  Phys.\ Rev.\ D {\bf 62} (2000) 093023.

\bibitem{Chen:2001qf}
  S.~Chen, M.~Davier, E.~Gamiz, A.~Hocker, A.~Pich and J.~Prades,
  Eur.\ Phys.\ J.\ C {\bf 22} (2001) 31.

\bibitem{Gamiz:2002nu}
  E.~G\'amiz, M.~Jamin, A.~Pich, J.~Prades and F.~Schwab,
  JHEP {\bf 0301} (2003) 060.

\bibitem{Gamiz:2004ar}
  E.~G\'amiz, M.~Jamin, A.~Pich, J.~Prades and F.~Schwab,
  Phys.\ Rev.\ Lett.\  {\bf 94} (2005) 011803.

\bibitem{Baikov:2004tk}
  P.~A.~Baikov, K.~G.~Chetyrkin and J.~H.~Kuhn,
  Phys.\ Rev.\ Lett.\  {\bf 95} (2005) 012003.

\bibitem{Gamiz:2007qs}
  E.~G\'amiz, M.~Jamin, A.~Pich, J.~Prades and F.~Schwab,
  PoS KAON {\bf } (2008) 008.

\bibitem{Aubert:2007jh}
  B.~Aubert {\it et al.}  [BaBar Collaboration],
  Phys.\ Rev.\ D {\bf 76} (2007) 051104.

\bibitem{Epifanov:2007rf}
  D.~Epifanov {\it et al.}  [Belle Collaboration],
  Phys.\ Lett.\ B {\bf 654} (2007) 65.

\bibitem{Jamin:2006tk}
  M.~Jamin, A.~Pich and J.~Portol\'es,
  Phys.\ Lett.\ B {\bf 640} (2006) 176.

\bibitem{Moussallam:2007qc}
  B.~Moussallam,
  Eur.\ Phys.\ J.\ C {\bf 53} (2008) 401.

\bibitem{Jamin:2008qg}
  M.~Jamin, A.~Pich and J.~Portol\'es,
  Phys.\ Lett.\ B {\bf 664} (2008) 78.

\bibitem{Boito:2008fq}
  D.~R.~Boito, R.~Escribano and M.~Jamin,
  Eur.\ Phys.\ J.\ C {\bf 59} (2009) 821.

\bibitem{Boito:2010me}
  D.~R.~Boito, R.~Escribano and M.~Jamin,
  JHEP {\bf 1009} (2010) 031.

\bibitem{GomezDumm:2003ku}
 D.~G\'omez Dumm, A.~Pich and J.~Portol\'es,
 Phys.\ Rev.\ D {\bf 69} (2004) 073002.

\bibitem{Dumm:2009kj}
  D.~G.~Dumm, P.~Roig, A.~Pich and J.~Portol\'es,
  Phys.\ Rev.\ D {\bf 81} (2010) 034031.

\bibitem{Dumm:2009va}
  D.~G.~Dumm, P.~Roig, A.~Pich and J.~Portol\'es,
  Phys.\ Lett.\ B {\bf 685} (2010) 158.

\bibitem{Dumm:2012vb}
  D.~G.~Dumm and P.~Roig,
  Phys.\ Rev.\ D {\bf 86} (2012) 076009.

 \bibitem{Guo:2010dv}
  Z.~-H.~Guo and P.~Roig,
  Phys.\ Rev.\ D {\bf 82} (2010) 113016.

\bibitem{Roig:2013??}
P.~Roig, A.~Guevara and G.~L\'opez Castro,
[arXiv:1306.1732 [hep-ph]], to be published in Phys.\ Rev.\ D.

\bibitem{Bartelt:1996iv}
  J.~E.~Bartelt {\it et al.}  [CLEO Collaboration],
  Phys.\ Rev.\ Lett.\  {\bf 76} (1996) 4119.

\bibitem{Buskulic:1996qs}
  D.~Buskulic {\it et al.}  [ALEPH Collaboration],
  Z.\ Phys.\ C {\bf 74} (1997) 263.

\bibitem{Inami:2008ar}
  K.~Inami {\it et al.}  [Belle Collaboration],
  Phys.\ Lett.\ B {\bf 672} (2009) 209.

\bibitem{delAmoSanchez:2010pc}
  P.~del Amo Sanchez {\it et al.}  [BaBar Collaboration],
  Phys.\ Rev.\ D {\bf 83} (2011) 032002.

\bibitem{Beringer:1900zz}
  J.~Beringer {\it et al.}  [Particle Data Group Collaboration],
  Phys.\ Rev.\ D {\bf 86} (2012) 010001.

\bibitem{Lees:2012ks}
  J.~P.~Lees {\it et al.}  [BaBar Collaboration],
  Phys.\ Rev.\ D {\bf 86} (2012) 092010.

\bibitem{Pich:1987qq}
  A.~Pich,
  Phys.\ Lett.\ B {\bf 196} (1987) 561.

\bibitem{Braaten:1989zn}
  E.~Braaten, R.~J.~Oakes and S.~-M.~Tse,
  Int.\ J.\ Mod.\ Phys.\ A {\bf 5} (1990) 2737.

\bibitem{Li:1996md}
  B.~A.~Li,
  Phys.\ Rev.\ D {\bf 55} (1997) 1436.

\bibitem{Aubrecht:1981cr}
  G.~J.~Aubrecht, II, N.~Chahrouri and K.~Slanec,
  Phys.\ Rev.\ D {\bf 24} (1981) 1318.

\bibitem{Actis:2010gg}
 S.~Actis {\it et al.},
 Eur.\ Phys.\ J.\  C {\bf 66} (2010) 585.

\bibitem{Kimura:2012}
D.~Kimura, K.~Y.~Lee, T.~Morozumi,
Prog.\  Theor.\  Exp.\  Phys.\  {\bf 2013} (2013) 053803.

\bibitem{Jadach:1990mz}
  S.~Jadach, J.~H.~Kuhn and Z.~Was,
  Comput.\ Phys.\ Commun.\  {\bf 64} (1990) 275.

\bibitem{Jadach:1993hs}
  S.~Jadach, Z.~Was, R.~Decker and J.~H.~Kuhn,
  Comput.\ Phys.\ Commun.\  {\bf 76} (1993) 361.

\bibitem{Shekhovtsova:2012ra}
  O.~Shekhovtsova, T.~Przedzinski, P.~Roig and Z.~Was,
  Phys.\ Rev.\ D {\bf 86} (2012) 113008

\bibitem{Nugent:2013hxa}
I.~M.~Nugent, T.~Przedzinski, P.~Roig, O.~Shekhovtsova and Z.~Was,
  arXiv:1310.1053 [hep-ph].

\bibitem{Weinberg:1978kz}
  S.~Weinberg,
  Physica A {\bf 96} (1979) 327.

\bibitem{Gasser:1983yg}
  J.~Gasser and H.~Leutwyler,
  Annals Phys.\  {\bf 158} (1984) 142.

\bibitem{Gasser:1984gg}
  J.~Gasser and H.~Leutwyler,
  Nucl.\ Phys.\ B {\bf 250} (1985) 465.

\bibitem{Ecker:1988te}
  G.~Ecker, J.~Gasser, A.~Pich and E.~de Rafael,
  Nucl.\ Phys.\ B {\bf 321} (1989) 311.

\bibitem{Ecker:1989yg}
  G.~Ecker, J.~Gasser, H.~Leutwyler, A.~Pich and E.~de Rafael,
  Phys.\ Lett.\ B {\bf 223} (1989) 425.

\bibitem{Gasser:1984ux}
  J.~Gasser and H.~Leutwyler,
  Nucl.\ Phys.\ B {\bf 250} (1985) 517.

\bibitem{Erler:2002mv}
  J.~Erler,
  Rev.\ Mex.\ Fis.\  {\bf 50} (2004) 200.

\bibitem{Ambrosino:2006gk}
  F.~Ambrosino {\it et al.}  [KLOE Collaboration],
  Phys.\ Lett.\ B {\bf 648} (2007) 267.

\bibitem{Kaiser:1998ds}
  R.~Kaiser and H.~Leutwyler,
  In *Adelaide 1998, Nonperturbative methods in quantum field theory* 15-29.

\bibitem{Kaiser:2000gs}
  R.~Kaiser and H.~Leutwyler,
  Eur.\ Phys.\ J.\ C {\bf 17} (2000) 623.

\bibitem{'tHooft:1973jz}
  G.~'t Hooft,
  Nucl.\ Phys.\ B {\bf 72} (1974) 461.

\bibitem{'tHooft:1974hx}
  G.~'t Hooft,
  Nucl.\ Phys.\ B {\bf 75} (1974) 461.

\bibitem{Witten:1979kh}
  E.~Witten,
  Nucl.\ Phys.\ B {\bf 160} (1979) 57.

\bibitem{Antonelli:2010yf}
  M.~Antonelli, V.~Cirigliano, G.~Isidori, F.~Mescia, 
 {\it et al.},
  Eur.\ Phys.\ J.\ C {\bf 69} (2010) 399.

\bibitem{Bijnens:1999sh}
  J.~Bijnens, G.~Colangelo and G.~Ecker,
  JHEP {\bf 9902} (1999) 020,

\bibitem{Bijnens:1999hw}
 J.~Bijnens, G.~Colangelo and G.~Ecker,
  Annals Phys.\  {\bf 280} (2000) 100.

\bibitem{Bijnens:2001bb}
  J.~Bijnens, L.~Girlanda and P.~Talavera,
  Eur.\ Phys.\ J.\ C {\bf 23} (2002) 539.

\bibitem{Manohar:1998xv}
A.~V.~Manohar,
 Published in 'Les Houches 1997, Probing the standard model of particle interactions, Pt. 2' 1091-1169.

\bibitem{Pich:2002xy}
 A.~Pich,
 Published in 'Tempe 2002, Phenomenology of large $N_C$ $QCD$' 239-258.

\bibitem{Cirigliano:2003yq}
  V.~Cirigliano, G.~Ecker, H.~Neufeld and A.~Pich,
  JHEP {\bf 0306} (2003) 012.

\bibitem{Jamin:2000wn}
  M.~Jamin, J.~A.~Oller and A.~Pich,
  Nucl.\ Phys.\ B {\bf 587} (2000) 331.

\bibitem{Jamin:2001zq}
  M.~Jamin, J.~A.~Oller and A.~Pich,
  Nucl.\ Phys.\ B {\bf 622} (2002) 279.

\bibitem{Jamin:2001zr}
  M.~Jamin, J.~A.~Oller and A.~Pich,
  Eur.\ Phys.\ J.\ C {\bf 24} (2002) 237.

\bibitem{Jamin:2006tj}
  M.~Jamin, J.~A.~Oller and A.~Pich,
  Phys.\ Rev.\ D {\bf 74} (2006) 074009.

\bibitem{Bernard:1990kw}
  V.~Bernard, N.~Kaiser and U.~G.~Meissner,
  Nucl.\ Phys.\ B {\bf 357} (1991) 129.

\bibitem{SanzCillero:2002bs}
  J.~J.~Sanz-Cillero and A.~Pich,
  Eur.\ Phys.\ J.\ C {\bf 27} (2003) 587.

\bibitem{Mateu:2007tr}
 V.~Mateu and J.~Portol\'es,
 Eur.\ Phys.\ J.\  C {\bf 52} (2007) 325.

\bibitem{RuizFemenia:2003hm}
 P.~D.~Ruiz-Femen\'{\i}a, A.~Pich and J.~Portol\'es,
 JHEP {\bf 0307} (2003) 003.

\bibitem{Cirigliano:2004ue}
 V.~Cirigliano, G.~Ecker, M.~Eidem\"uller, A.~Pich and J.~Portol\'es,
 Phys.\ Lett.\ B {\bf 596} (2004) 96.

\bibitem{Cirigliano:2005xn}
 V.~Cirigliano, G.~Ecker, M.~Eidem\"uller, R.~Kaiser, A.~Pich and J.~Portol\'es,
 JHEP {\bf 0504} (2005) 006.

\bibitem{Cirigliano:2006hb}
 V.~Cirigliano, G.~Ecker, M.~Eidem\"uller, R.~Kaiser, A.~Pich and J.~Portol\'es,
 Nucl.\ Phys.\  B {\bf 753} (2006) 139.

\bibitem{Kampf:2011ty}
  K.~Kampf and J.~Novotny,
  Phys.\ Rev.\ D {\bf 84} (2011) 014036.

\bibitem{Masjuan:2007ay}
  P.~Masjuan and S.~Peris,
  JHEP {\bf 0705} (2007) 040.

\bibitem{Lepage:1979zb}
  G.~P.~Lepage and S.~J.~Brodsky,
  Phys.\ Lett.\ B {\bf 87} (1979) 359.

\bibitem{Lepage:1980fj}
  G.~P.~Lepage and S.~J.~Brodsky,
  Phys.\ Rev.\ D {\bf 22} (1980) 2157.

\bibitem{GomezDumm:2000fz}
  D.~G\'omez Dumm, A.~Pich and J.~Portol\'es,
  Phys.\ Rev.\  D {\bf 62} (2000) 054014.

\bibitem{Pich:2001pj}
  A.~Pich, J.~Portol\'es,
  Phys.\ Rev.\  {\bf D63 } (2001)  093005.

\bibitem{Dumm:2013zh}
  D.~G.~Dumm and P.~Roig,
  arXiv:1301.6973 [hep-ph] and work in progress.

\bibitem{Guerrero:1997ku}
  F.~Guerrero and A.~Pich,
  Phys.\ Lett.\  B {\bf 412} (1997) 382.

\bibitem{Guerrero:1998hd}
  F.~Guerrero,
  Phys.\ Rev.\ D {\bf 57} (1998) 4136.

\bibitem{Guo:2012ym}
  Z.~-H.~Guo, J.~A.~Oller and J.~Ruiz de Elvira,
  Phys.\ Lett.\ B {\bf 712} (2012) 407.

\bibitem{Guo:2012yt}
  Z.~-H.~Guo, J.~A.~Oller and J.~Ruiz de Elvira,
 Phys.\ Rev.\ D {\bf 86} (2012) 054006

\bibitem{Guo:2011pa}
  Z.~-H.~Guo and J.~A.~Oller,
  Phys.\ Rev.\ D {\bf 84} (2011) 034005.

\bibitem{GomezNicola:2001as}
  A.~G\'omez Nicola and J.~R.~Pel\'aez,
  Phys.\ Rev.\ D {\bf 65} (2002) 054009.

\bibitem{Escribano:2010wt}
 R.~Escribano, P.~Masjuan and J.~J.~Sanz-Cillero,
 JHEP {\bf 1105} (2011) 094.

\bibitem{Escribano:2002iv}
  R.~Escribano, A.~Gallegos, J.~L.~Lucio M, G.~Moreno and J.~Pestieau,
  Eur.\ Phys.\ J.\ C {\bf 28} (2003) 107.

\bibitem{Pallante:2000hk}
  E.~Pallante and A.~Pich,
  Nucl.\ Phys.\ B {\bf 592} (2001) 294.

\bibitem{Oller:2000ma}
  J.~A.~Oller, E.~Oset and A.~Ramos,
  Prog.\ Part.\ Nucl.\ Phys.\  {\bf 45} (2000) 157.
\end{thebibliography}
\end{document}